\documentclass[final,5p,times,twocolumn,number]{elsarticle}

\usepackage{amssymb}
\usepackage{lipsum}
\usepackage{amsmath,amsfonts}
\usepackage{amsthm}
\usepackage{amssymb}
\usepackage{algorithmic}
\usepackage{algorithm}
\usepackage{array}
\usepackage{mathrsfs}
\usepackage{textcomp}
\usepackage{stfloats}
\usepackage{url}
\usepackage{amsmath}
\usepackage{mathtools}
\usepackage{verbatim}
\usepackage{graphicx}
\usepackage{autobreak}
\usepackage{nicematrix}
\usepackage{arydshln}
\usepackage{latexsym}
\usepackage{booktabs}
\usepackage{nomencl}
\usepackage{graphicx}

\allowdisplaybreaks[4]

\makeatletter
\def\ps@pprintTitle{%
	\let\@oddhead\@empty
	\let\@evenhead\@empty
	\let\@oddfoot\@empty
	\let\@evenfoot\@empty
}
\def\pprintMaketitle{
	\clearpage
	\thispagestyle{pprintTitle}%
	\gdef\@fpheader{}
	\gdef\@journal{}%
	\let\@oddhead\@empty
	\let\@evenhead\@empty
	\maketitle
}
\makeatother

\begin{document}

\begin{frontmatter}



\title{Online Distributed Optimization for Spatio-temporally Constrained P2P Energy Trading}


\author[first]{Junhong Liu}
\author[first]{Qinfei Long}
\author[first,second]{Rong-Peng Liu}
\author[first]{Wenjie Liu}
\author[first]{Yunhe Hou}
\affiliation[first]{organization={Department of Electrical and Electronic Engineering, The University of Hong Kong},
            city={Hong Kong SAR},
            postcode={999077},
            country={China}}
\affiliation[second]{organization={Department of Electrical and Computer Engineering, McGill University},
	city={Montréal},
	postcode={QC H3A 0E9},
	country={Canada}}     
\cortext[cor1]{Corresponding author: Yunhe Hou (yhhou@eee.hku.hk)}	       
\begin{abstract}
The proliferation of distributed renewable energy triggers the peer-to-peer (P2P) energy market formations. To make profits, prosumers equipped with photovoltaic (PV) panels and even the energy storage system (ESS) can actively participate in the real-time P2P energy market and trade energy. However, in real situations, system states such as energy demands and renewable energy power generation are highly uncertain, making it difficult for prosumers to make optimal real-time decisions. Moreover, severe problems with the physical network can arise from the real-time P2P energy trading, such as bus voltage violations and line overload. To handle these problems, this work first formulates the real-time P2P energy trading problem as a spatio-temporally constrained stochastic optimization problem by considering ESS and the spatial physical network constraints. To deal with the uncertainties online, a modified Lyapunov optimization method is innovatively proposed to approximately reformulate the stochastic optimization problem into an online one by relaxing the time-coupling constraints. Compared with the state-of-the-art online methods, the proposed one renders more flexibility and better performance for the real-time P2P energy market operation. Additionally, to protect the prosumers’ privacy, an online distributed algorithm based on the consensus alternating direction method of multipliers (ADMM) is developed to solve the reformulated online problem by decoupling the spatial constraints. The theoretical near-optimal performance guarantee of the proposed online distributed algorithm is derived, and its performance can be further improved by minimizing the performance gap. Simulation results demonstrate that the proposed online distributed algorithm can guarantee the fast, stable, and safe long-term operation of the real-time P2P energy market. Allied with a performance gap minimization step, the proposed online distributed algorithm can obtain approximately a 53.10\% cost reduction compared to the greedy algorithm for the intraday market operation.
\end{abstract}




\end{frontmatter}




\section{Introduction}
\label{introduction}
The rapid augment of the latest renewable energy power generation equipment, such as distributed rooftop PV panels, has paved the way for new energy markets. One of the promising markets is the P2P energy trading market, which aims to achieve energy self-consumption in local communities based on the resource-sharing concept \cite{morstyn2018using}. All the stakeholders could benefit from this market. Specifically, prosumers could obtain higher welfare than merely trading with the utility company via the traditional feed-in-tariff (FIT) scheme. The utility company could make profits from the reduced peak demands through the P2P energy trading \cite{zhang2018peer}. Since system states, such as the renewable energy power generation, energy demands, and price signals from the utility company, are highly uncertain, actions need to be performed by the prosumers as fast as possible to balance the energy power generation and demands. Thus, it requires the P2P energy market to be operated in a real-time manner \cite{liu2017distributed}. Meanwhile, to take advantage of the renewable energy resources and promote the P2P energy market, ESS could be helpful auxiliary equipment \cite{aminlou2022peer}. ESS is a kind of time-coupling entities, which could help reduce the variations of intermittent renewable energy power generation and uncertain demands \cite{huangoptimization,nguyen2018optimizing}. In \cite{nguyen2018optimizing}, up to 14\% higher savings are achieved by households through the coordination of PV systems and ESS than by only incorporating PV systems in the P2P energy market. Especially with the prices of ESS decreasing each year (e.g., The price of the Lithium-ion based battery pack fell from 1100 USD/kWh in 2010 to 156 USD/kWh in 2019), the deployment of ESS in the power system will become more common in the near future \cite{hannan2021battery}. Therefore, it is meaningful to consider ESS for the real-time P2P energy market operation.

Nevertheless, high penetrations of uncertain renewable energy power generation into the power system could cause potential problems such as voltage violations and line overload \cite{song2017security,pillai2021facilitating}. Especially when the real-time P2P energy market is in operation, energy flows over the electric wires are more intricate, and problems with the physical network become progressively more severe if no appropriate measures are taken \cite{leong2019auction}. As a result, physical network constraints must be considered for the operation of a real-time P2P energy market. Tradition-ally, the P2P energy market could be operated in a centralized manner \cite{zhong2020cooperative,morstyn2018bilateral}. By contrast, a distributed manner is preferred for the real-time P2P energy market operation. To begin with, a distributed manner could meet the requirements of the real-time P2P energy market for fast market clearing. The distributed manner could alleviate the computational burden once placed on a single node by the centralized manner and thus accelerate clearing the energy trading problem \cite{peng2016distributed}. Besides, a distributed manner could preserve the sensitive information, including private parameters of the utility functions that quantity agents’ preferences, which also stimulates prosumers to actively participate in the market \cite{paudel2020peer}. However, when considering the network constraints as well as the P2P energy constraints, prosumers become spatially correlated, which conflicts with the distributed operation manner of the real-time P2P energy market. Moreover, incorporating time-coupling entities such as ESS into the real-time P2P energy market adds new challenges for prosumers in terms of making optimal real-time decisions and simultaneously guaranteeing the performance for the market operation. This work mainly focus-es on the two problems and proposes an online distributed optimization method for real-time energy trading in the P2P energy market.

The state-of-the-art literature on the P2P energy market could be classified into two categories based on the existence of time-coupling entities: the single-time-slot P2P energy trading problem and the multi-time-slot P2P energy trading problem. In the first category, researchers ignore the impacts of time-coupling entities and mainly focus on how to design a fully decentralized trading mechanism for the P2P energy market \cite{paudel2020peer2}, how to consider the physical network constraints in the market \cite{kim2019p2p}, how to devise a faster distributed trading algorithm \cite{feng2022peer}, or how to design the communication-efficient trading scheme \cite{umer2021novel}. Though meaningful, these results can-not be easily extended to the multi-time-slot P2P energy trading problem. In the presence of time-coupling entities, if optimal decisions are still made at each time slot for the multi-time-slot optimization problem, also referred to as the greedy algorithm in this paper, the system may obtain the suboptimal performance over the whole operation period \cite{shi2015real}. More specifically, if being unaware of the future lower buying prices of electricity from the utility company, prosumers equipped with ESS would fully charge the battery at the current high buying price, since it might be the optimal decision at the current time slot. When the buying price goes down, prosumers have no ability to store the relatively cheaper electricity and the suboptimal performance emerges.

On the contrary, the second category of the state-of-the-art literature considers time-coupling entities in the P2P energy trading problem. In this category, one typical way to deal with the suboptimal performance is to formulate the problems as the day-ahead scheduling problem \cite{paudel2020peer,iqbal2021novel,jia2022security,li2018distributed,paudel2018peer,zhang2019novel,khorasany2019decentralized}. For the day-ahead market operation period, day-ahead system states, such as renewable energy power generation, energy demands, and price signals from the utility company, are all assumed to be known or precisely predicted. For instance, in \cite{li2018distributed}, the day-ahead PV power generation and uncontrollable load data are assumed to be perfectly forecasted, and the authors aim to solve the multi-time-slot P2P energy trading problem through the blockchain-based framework when ESS is present. Besides, in \cite{paudel2018peer}, the day-ahead renewable energy power generation and load data of each prosumer equipped with ESS are also known in advance, and the problem is formulated by the hybrid game-theoretic models to schedule ESS in the P2P energy market. However, it is impractical and unrealistic to obtain the perfect information of day-ahead and even week-ahead system states. First, in order to make predictions through the data-driven methods, such as the deep learning or federated learning methods \cite{tina2021state,bouachir2022federatedgrids,chen2019realistic}, massive historical data of the system states should be collected and stored. Each prosumer is also required to be equipped with an additional high-performance computing unit in order to train and update the complex predictive models. It is not economical and practical for household prosumers. Furthermore, system states, such as the renewable energy power generation and uncontrollable loads from electrical vehicle owners, are random and highly uncertain. It is hard to make the exactly perfect predictions through the historical data, and the ac-curacy of predictions also highly depends on the selection of features and the quality of the collected data \cite{tina2021state,zhang2022application,shahriar2020machine,wang2019lyapunov,qin2014modeling}. Specifically, for the P2P energy market, energy demands and generation need to be balanced locally in a real-time manner, and the length of the time slot could be 15 minutes or even less \cite{guo2021online}. To avoid relying on the cumbersome data-driven methods to solve the multi-time-slot problem and simultaneously ensure system performance, a more lightweight and appropriate optimization method is necessary for the real-time P2P energy trading.

To deal with the drawbacks above, this paper considers the explicit network constraints and time-coupling entity in the real-time P2P energy trading problem, and formulates it as a spatial-temporally constrained stochastic optimization problem. Then we propose a modified Lyapunov optimization method to relax the time-coupling constraints. Accordingly, the stochastic optimization problem is approximately reformulated into an online one. The proposed method permits more choices of feasible solutions for the operating variables with-in the boundary constraints. The enlarged solution space thus leads to more flexibility and better performance for the market operation. Moreover, to protect the prosumers’ privacy, a consensus ADMM is proposed to solve the online optimization problem in a distributed manner. To improve the computational efficiency for clearing the real-time P2P energy trading problem, we further derive the closed-form solutions for all the decomposed sub-problems. The theoretical near-optimal performance guarantee of the proposed online distributed algorithm is derived, and the proposed algorithm’s performance could be further improved by minimizing the performance gap. Considering differences of this work from the existing literature, this paper contributes to existing researches in the following aspects:

\begin{itemize}
	\item{The real-time P2P energy trading problem is formulated as a spatial-temporally constrained stochastic optimization problem by considering the time-coupling entity and explicit physical network constraints.}
	
	\item{A modified Lyapunov optimization method is proposed to approximately reformulate the stochastic optimization problem into an online one, which owns better flexibility and performance for the real-time P2P energy market operation compared with the state-of-the-art methods.}
	
	\item{For protecting the prosumers’ privacy, an online distributed algorithm based on the consensus ADMM is developed to solve the reformulated online P2P energy trading problem, and the closed-form solutions for all the decomposed sub-problems are derived to significantly increase the computational efficiency.}
	
	\item{The theoretical near-optimal performance guarantee of the proposed online distributed algorithm is derived. Allied with a performance gap minimization step, the proposed online distributed algorithm obtains better performance for the real-time P2P energy trading.}
\end{itemize}
The rest of this paper is organized as follows: Section 2.2 presents the formulation for the real-time P2P energy trading problem; Section 2.3 describes the proposed online optimization method; Section 2.4 includes the pro-posed online distributed algorithm based on the consensus ADMM and the closed-form solutions for all the decomposed sub-problems; Section 2.5 consists of the online distributed algorithm’s performance guarantee; Section 2.6 provides the experiment setups and discussions of 7 case studies. Finally, conclusions for this paper are summarized in Section 2.7.

\section{Real-Time P2P Energy Trading Problem}
\begin{figure*}[!hbp]
	\centering
	\includegraphics[width=0.65\linewidth]{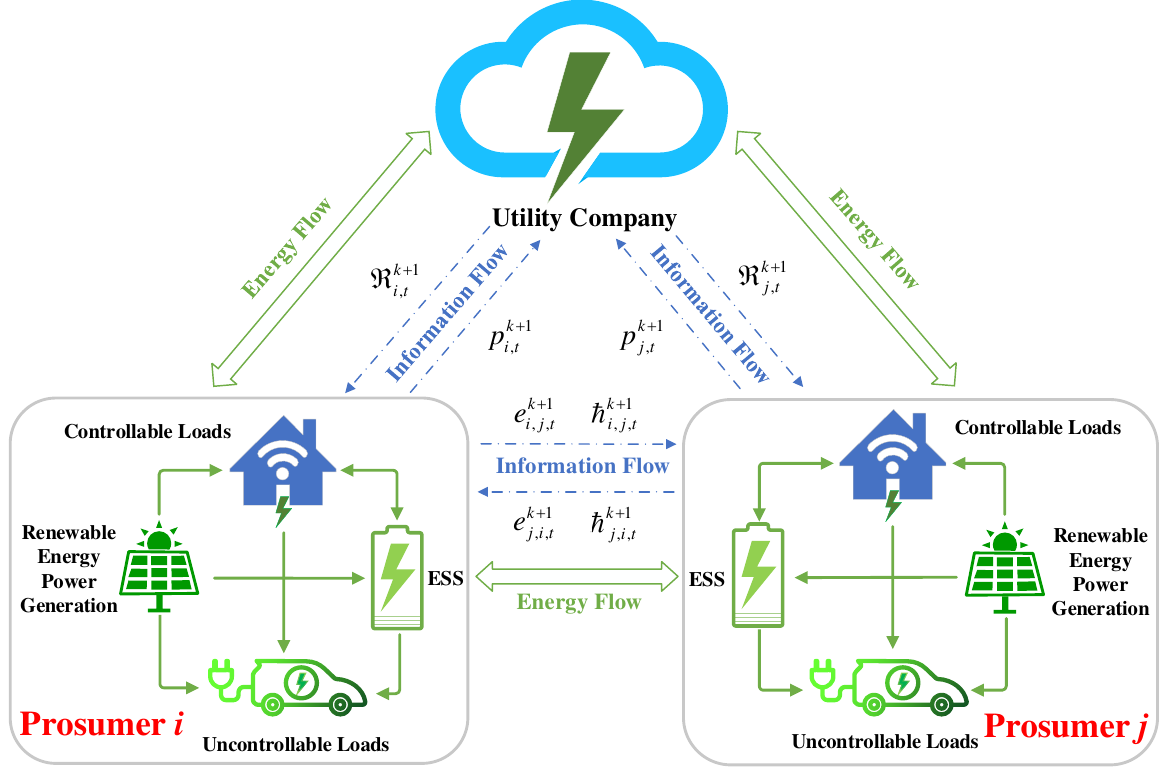}
	\caption{Scheme of the real-time P2P energy trading scenario.}
	\label{apen_1}
\end{figure*}

Relevant constraints and models for the real-time P2P energy trading are delineated in this section. The real-time P2P energy market is operated under the distribution network, where each prosumer is located at each bus. The utility company is introduced to balance the overall power flow among different entities. Specifically, the reformulated spatial linearized DistFlow constraints are employed to ensure that the traded power flow is within the secure regions. ESS and the bilateral P2P trading model impose additional time-coupling and spatial constraints. The spatial-temporally constrained stochastic optimization problem for the real-time P2P energy trading is established to minimize the time-averaged costs over the operation period, which is equivalent to maximizing the time-averaged social warfare. Prosumers can interact with each other and the utility company through both information flow and energy flow, and the scheme of a real-time P2P energy trading scenario is denoted as Figure \ref{apen_1}. Each prosumer is required to share parts of its decision variables with other agents through the information flow, and detailed information exchange procedures are delineated in Figure \ref{apen_2} and Algorithm \ref{apen_a1}.

\subsection{System Model and Constraints}
1) Network Constraints: Consider a radial distribution network. It is denoted through a directed graph. Let $\mathcal{N}:=\{0,1,...,N\}$ is the set for buses and $|\mathcal{L}|=(N+1)-1=N$ is the number of line segments connected by buses. Let $\mathcal{N}_p=\mathcal{N}/\{0\}$ denote the set of buses except for the substation bus 0. The power flow model at time slot $t$ is linearized (LinDistFlow) in the following form for a given line segment $\ell: (i,j):$:
\begin{subequations}
	\begin{align}
		& P_{\ell,t}-\sum_{m \in C_j}(P_{j,m,t})+p_{j,t}=0 \label{apen_f1}\\
		& Q_{\ell,t}-\sum_{m \in C_j}(Q_{j,m,t})+q_{j,t}=0 \label{apen_f2}\\
		& v_{i,t}-v_{j,t}=2r_{i,j,t}P_{\ell,t}+2x_{i,j,t}Q_{\ell,t}, \label{apen_f3} 
	\end{align}
\end{subequations} 
where $C_j$ is the set of children buses of bus $j$ and $v_{j,t}$ is the square of voltage magnitude at the bus $j$ at time slot $t$. The linear approximation error is estimated to be the order of 1\% \cite{zhou2019hierarchical}.  
Furthermore, the above linearized model is reformulated by a graph incidence matrix, $\mathcal{H}$, with the size of $N \times N$\cite{zhu2015fast,ullah2021peer}. The $\ell-$throw of the matrix $\mathcal{H}$ represents the line $\ell$ connected by the starting bus $i$ and ending bus $j$, with $\mathcal{H}_{\ell,i}=1$ and $\mathcal{H}_{\ell,j}=-1$. The reformulated model is summarized as follows:
\begin{subequations}
	\begin{align}
		 \mathbf{P}=\mathcal{H}^{-T} \mathbf{p}& \label{apen_f4}\\
	     \mathbf{Q}=\mathcal{H}^{-T} \mathbf{q}& \label{apen_f5}\\
	     \mathbf{v}=v_0+2\mathcal{H}^{-1}D_r\mathcal{H}^{-T}\mathbf{p}&+2\mathcal{H}^{-1}D_x\mathcal{H}^{-T}\mathbf{q}, \label{apen_f6}
	\end{align}
\end{subequations} 
where $v_0$ is square of the per-unit voltage at the substation bus 0, and $v_0=1.0^2$. The bold symbols, $\mathbf{P}, \mathbf{Q}, \mathbf{p}, \mathbf{q}$ and $\mathbf{v}$, represent vectors containing corresponding variables. $D_r$ and $D_x$ are the diagonal matrices with each diagonal term to be $r_{i,j}$ and $x_{i,j}$, respectively. $\mathcal{H}^{-T}$ is the transposed matrix of $\mathcal{H}^{-1}$.
Meanwhile, the network constraints are given as follows in terms of the active and reactive power flow at the line segment $\ell:(i,j)$, and the voltage of the bus $j$:
\begin{subequations}
	\begin{align}
		& P_{\ell}^{min} \le P_{\ell,t} \le P_{\ell}^{max} \label{apen_f7}\\
		& Q_{\ell}^{min} \le Q_{\ell,t} \le Q_{\ell}^{max} \label{apen_f8}\\
		& v_{j}^{min} \le v_{j,t} \le v_{j}^{max}, \label{apen_f9}
	\end{align}
\end{subequations} 
where $v_j^{min}$ and $v_j^{max}$ are the minimum and maximum allowed voltage magnitude square at the bus $j$. A similar voltage constraint also works for the other bus $i$ at the line segment $\ell:(i,j)$.

\textbf{Remark 2.1:} The physical network topology is assumed to be unchanged during the online peer-to-peer energy market operation, where the graph incidence matrix $\mathcal{H}$, the line resistance matrix $D_r$ and reactance matrix $D_x$ remain constant. Meanwhile, the reactive power injection at bus $j$, $q_{j,t}$, is assumed to be in a constant proportion to the active power injection $p_{j,t}$ over the online optimization period, which means $q_{j,t}=\mathcal{X}_j \cdot p_{j,t}$ \cite{ullah2021peer}. This assumption on the constant proportion can be lifted in order to better provide the reactive power support service, and one more variable, $q_{j,t}$, should be transmitted by each prosumer to the utility company during the online optimization period.

2) Energy Storage System: Let $S_{i,t}$ represent the State of Charge (SoC) of ESS of prosumer $i$ at time slot $t$ and let $w_{i,t}$ denote the corresponding charging or discharging energy at the current time slot. Let $\kappa_i \in (0,1]$ to be the charging/discharging coefficient that models the energy loss over time, which is treated as a constant for each prosumer. The dynamics of ESS from prosumer $i$ over time is denoted as: 
\begin{subequations}
	\begin{align}
		& S_{i,t+1}=\kappa_i S_{i,t}+w_{i,t} \label{apen_f10}\\
		& S_{i}^{min} \le S_{i,t} \le S_{i}^{max} \label{apen_f11}\\
		& w_{i}^{min} \le w_{i,t} \le w_{i}^{max}, \label{apen_f12}
	\end{align}
\end{subequations} 
where $w_i^{min} \le 0$ and $w_i^{max} \ge 0$ represent the maximum allowed discharging and charging energy in one time slot. $S_i^{min}$ and $S_i^{max}$ represent the minimum and maximum allowed values of SoC of ESS.

\textbf{Remark 2.2}: Constraint \eqref{apen_f10} from the dynamics of ESS is time-coupling, which challenges the online operation of the real-time P2P energy market. The load demand $d_{i,t}$ that accounts for the electrical vehicles can be the time-coupling variable, and can be regarded as the subcase of ESS. However, considering the potentially random behaviors of electrical vehicles, it is further simplified here and considered as the real-time load.

3) Net Demand: Let $g_{i,t}$ be the renewable energy power generated from PV panels owned by prosumer  $i$ at time slot $t$. Correspondingly, let $d_{i,t}$ represent its aggregated demands containing the controllable and uncontrollable loads. Assume the controllable load is flexible and can be rescheduled. The net energy power generation is described as below:
\begin{subequations}
	\begin{align}
		& p_{i,t}=g_{i,t}-d_{i,t} \label{apen_f13}\\
		& d_{i,t}^{min} \le d_{i,t} \le d_{i,t}^{max}. \label{apen_f14}
	\end{align}
\end{subequations} 

4) Bilateral P2P Trading Constraints: Consider that the P2P energy market consists of a set of buyers $\mathcal{N}_b \subseteq \mathcal{N}_p$ and sellers $\mathcal{N}_s \subseteq \mathcal{N}_p$, where $\mathcal{N}_s \cap \mathcal{N}_b = \emptyset$. Depending on its net energy power generation at the current time slot $p_{i,t}$, each prosumer reports to the system to be either a buyer or a seller. Namely, if the net energy power generation of one prosumer is positive, then this prosumer becomes a seller, and vice visa. Define $\mathcal{N}_i$ to be the set of trading partners for the prosumer $i$. Specifically, if the prosumer $i$ is a buyer, then its trading partners belong to the set of sellers and $\mathcal{N}_i \subseteq \mathcal{N}_s$; conversely, if the prosumer $i$ is a seller, then $\mathcal{N}_i \subseteq \mathcal{N}_b$. Define $e_{i,j,t}$ to be the traded energy between prosumers $i$ and $j$ at time slot $t$. If $e_{i,j,t} \le 0$ and $e_{i,j,t} \ge 0$, prosumer $i$ is a buyer and prosumer $j$ is a seller. Thus, the spatial constraints for the couple of prosumers $i$ and $j$ are formulated as follows:
\begin{subequations}
	\begin{align}
		& e_{i,j,t} \le 0, e_{j,i,t} \ge 0, i \in \mathcal{N}_b \text{ and } j \in \mathcal{N}_i \label{apen_f15}\\
		& e_{i,j,t} \ge 0, e_{j,i,t} \le 0, i \in \mathcal{N}_s \text{ and } j \in \mathcal{N}_i \label{apen_f16}\\
		& e_{i,j,t}+e_{j,i,t}=0, j \in \mathcal{N}_i. \label{apen_f17}
	\end{align}
\end{subequations} 

5) Power Balance Constraints: To balance the power flow over the network and to satisfy demands of prosumers under different time slots when the renewable energy power generation from the PV panels is una-vailable, prosumers should also be able to interact with the utility company. Define $p_{i,t}^b$ and $p_{i,t}^s$ to be the energy that prosumer $i$ buys from the company and sells to the company at the current time slot, respectively. The energy balance for each prosumer $i$ at time slot $t$ is described as follows:
\begin{subequations}
	\begin{align}
		& p_{i,t}^b = [\sum_{j \in \mathcal{N}_s}e_{i,j,t}+w_{i,t}-p_{i,t}]^+, i \in \mathcal{N}_b \label{apen_f18}\\
		& p_{i,t}^s = [p_{i,t}-(\sum_{j \in \mathcal{N}_b}e_{i,j,t}+w_{i,t})]^+, i \in \mathcal{N}_s, \label{apen_f19} 
	\end{align}
\end{subequations} 
where $[\centerdot]^+$ represents the non-negative operator, which equals to $max \{\centerdot,0\}$.

\textbf{Remark 2.3}: If prosumer $i$ is a buyer, which means $p_{i,t}<0$, then it is assumed that this prosumer can only buy energy from the utility company at time slot $t$. Conversely, if this prosumer   is a seller, which means $p_{i,t}\ge 0$, this prosumer is assumed to only sell energy to the company at current time slot. 

6) Overall Objective Function: The objective function comprises four parts. They are the utility/cost for the prosumers trading energy in the P2P market, the cost of rescheduling loads, the cost of manipulating ESS, and the cost of interacting with the utility company. The quadratic function is employed to capture the prosumers’ satisfaction levels of sharing energy in the P2P energy market. Let $ \alpha_i \ge 0$ and $\beta_i \ge 0$  be coefficients of the cost/utility function. Besides, due to the involvement in the P2P energy market, prosumers’ actual demands might deviate from their preferred demand profiles. The quadratic function is introduced to model the discomfort cost of rescheduling the demand. Let $\gamma_i$ denote the coefficient for the discomfort cost function for the prosumer $i$. Manipulating ESS should shorten its service life. To avoid frequent operations of ESS, the cost function is introduced for ESS, which is modelled by the term $|\xi_i w_{i,t}|$. $\xi_i$ is the constant cost coefficient for prosumer $i$ to charge or discharge per unit of energy \cite{qin2015online}. Let $\lambda_t^s$ denote the price of selling per unit of energy to the utility company and $\lambda_t^b$ be the corresponding buying price from the utility company at time slot $t$. Thus, the overall objective function at time slot $t$ is represented by the following form:
\begin{align}
	& f_t(\mathbf{\Phi}_t)=\sum_{i \in \mathcal{N}_p} \{\alpha_i \sum_{j \in \mathcal{N}_i}e_{i,j,t}^2+\beta_i \sum_{j \in \mathcal{N}_i}e_{i,j,t}+\gamma_i(p_{i,t}-g_{i,t}+d_{i,t})^2 \notag\\
	& \quad\quad\quad +|\xi_iw_{i,t}|-\lambda_t^s \cdot p_{i,t}^s +\lambda_t^b \cdot p_{i,t}^b\}, \label{apen_f20}
\end{align}
where $\mathbf{\Phi}_t=\{\mathbf{e}_{1,t},p_{1,t},w_{1,t},...,\mathbf{e}_{N,t},p_{N,t},w_{N,t}\}$ and $\mathbf{e}_{1,t}=\{e_{1,1,t},...,e_{1,j,t}\}$ are the decision variables at time slot t. The system state at time slot $t$ is denoted by a collection of random variables as $\mathbf{R}_t=\{g_{i,1},d_{i,1},...,g_{N,1},d_{N,t}, \lambda_t^s,\lambda_t^b\}$, which includes the prosumers’ renewable energy power generation, prosumers’ energy demands, and price signals from the utility company. For the single-time-slot P2P energy trading problem, the decisions can be easily made according to the realization of the current system state. While considering multiple time slots, the future system states are uncertain and making the globally optimal decisions is challenging. Assuming all the mentioned system states are random variables over the operation period, we can formulate the P2P energy trading problem into a spatial-temporally constrained stochastic optimization problem.
\subsection{Real-Time P2P Energy Trading Optimization Problem}
At the beginning of the operation period, the future system states are uncertain. The objective is to minimize the expected time-averaged cost over the given operation period, which is formulated into the spatial-temporally constrained stochastic optimization problem \textbf{P1} as follows:
\begin{subequations}
	\begin{align}
		& \textbf{P1:}  \mathop{min}_{\Phi_t}   \mathop{lim}_{\mathcal{T} \rightarrow \infty} \frac{1}{\mathcal{T}} \sum_{t=1}^{t=\mathcal{T}} \{ \mathbb{E}[f_t(\mathbf{\Phi}_t)]\}  \\
		&\quad\quad s.t. \quad \eqref{apen_f4}-\eqref{apen_f19},
	\end{align}
\end{subequations}
where $\mathcal{T}$ is the total number of time slots for the given operation period. When the length of each time slot is small enough, such as 15 minutes or even 15 seconds, the P2P energy market can work in a real-time manner. Meanwhile, when the given operation period is as long as several weeks or even months, $\mathcal{T}$ is large enough that can be treated as an infinite value. It should be noted that prosumers need to cooperate with each other by sharing partial information in order to solve \textbf{P1} and to minimize overall total costs.
In \textbf{P1}, since constraint \eqref{apen_f10} is time-coupling, decisions of the current time slot should influence the ensuing decisions. Obtaining the globally optimal solutions of such a problem requires the detailed information of future system states, which is unrealistic. In the next section, we propose a modified Lyapunov optimization method to make online decisions for the P2P energy market without relying on the detailed future information, which has better flexibility than the state-of-the-art online methods and is guaranteed with the near-optimal performance.

\section{Online Optimization for Real-Time P2P Energy Trading}
Due to the intermittent renewable energy power generation, random energy demand, and volatile buying/selling prices from the utility company, \textbf{P1} denotes a spatial-temporally constrained stochastic optimization problem that is hard to be solved online. In this section, \textbf{P1} is reformulated into the online optimization problem \textbf{P2} by the proposed modified Lyapunov optimization method to relax the time-coupling constraints and to provide an enlarged solution space for the decision variables.

\subsection{Modified Lyapunov Optimization for Real-time P2P Energy Market}
By introducing the shifting variable $\epsilon_i$ based on the standard Lyapunov optimization theory \cite{zhong2019online,qin2015online,neely2022stochastic}, the virtual queue of the SoC of ESS at the time slot $t$ for prosumer $i$, $\tilde{S}_{i,t}$, is defined as below:
\begin{align}
	\tilde{S}_{i,t}=S_{i,t}+\epsilon_i.
\end{align}
Then the weighted quadratic Lyapunov function for the SoC of ESS, $L(\tilde{S}_{i,t})$, is formulated as:
\begin{align}
	L(\tilde{S}_{i,t})=\frac{1}{2}\delta_i \cdot \tilde{S}_{i,t}^2,
\end{align}
where $\delta_i$ is the wight parameter for each prosumer. Then the per-time-slot Lyapunov drift term, $\triangle(\tilde{S}_{i,t})$, at the time slot $t$ is obtained on the condition that $\tilde{S}_{i,t}$ is known:
\begin{align}
	\triangle(\tilde{S}_{i,t})=\mathbb{E}[L(\tilde{S}_{i,t+1})-L(\tilde{S}_{i,t})|\tilde{S}_{i,t}].
\end{align}
Based on the above defined per-time-slot Lyapunov drift term, we can obtain the following lemma. 

\textbf{Lemma 2.1}: Under any system states, the per-time-slot Lyapunov drift term has the following upper bound:
\begin{align}
	\triangle(\tilde{S}_{i,t}) \le M_i+\delta_i \cdot \mathbb{E}[\kappa_i \tilde{S}_{i,t}w_{i,t}+\epsilon_i(1-\kappa_i)w_{i,t}+\frac{w_{i,t}^2}{2}],
\end{align}
where $M_i$ is a constant for the given parameters $\{\delta_i,\epsilon_i,\kappa_i,S_i^{min},S_i^{max}\}$, and is denoted as below:
\begin{align}
	M_i=&\delta_i \cdot [\frac{1}{2}(\kappa_i^2-1)\cdot max\{(S_i^{max}+\epsilon_i),S_i^{min}+\epsilon_i)\} \notag \\
	&\qquad+\kappa_i(1-\kappa_i)\tilde{S}_{i,t}\epsilon_i+\frac{1}{2}(1-\kappa_i)^2\epsilon_i^2]. 
\end{align}
\begin{subequations}
\textit{Proof}:\begin{align}
	&\triangle(\tilde{S}_{i,t})=\frac{1}{2}\delta_i \cdot \mathbb{E}[(\kappa_i\tilde{S}_{i,t}+w_{i,t}+(1-\kappa_i)\epsilon_i)^2-\tilde{S}_{i,t}^2] \\
	&=\delta_i\cdot \mathbb{E}[\frac{1}{2}(\kappa_i^2-1)\tilde{S}_{i,t}^2+\kappa_i\tilde{S}_{i,t}w_{i,t}+\kappa_i(1-\kappa_i)\tilde{S}_{i,t}\epsilon_i \notag\\
	&\qquad +\frac{(w_{i,t}+(1-\kappa_i)\epsilon_i)^2}{2}] \\
	&=\delta_i\cdot \mathbb{E}[\frac{1}{2}(\kappa_i^2-1)\tilde{S}_{i,t}^2+\kappa_i(1-\kappa_i)\tilde{S}_{i,t}\epsilon_i+\frac{1}{2}(1-\kappa_i)^2\epsilon_i^2] \notag\\
	&\qquad +\delta_i\cdot \mathbb{E}[\frac{w_{i,t}^2}{2}+\kappa_i w_{i,t}\tilde{S}_{i,t}+(1-\kappa_i)w_{i,t}\epsilon_i] \\
	&=M_i+\delta_i\cdot\mathbb{E}[\frac{w_{i,t}^2}{2}+\kappa_iw_{i,t}\tilde{S}_{i,t}+(1-\kappa_i)w_{i,t}\epsilon_i],
\end{align}
\end{subequations}
where $M_i=\delta_i\cdot[0.5(\kappa_i^2-1)\cdot max\{(S_i^{min}+\epsilon_i)^2,(S_i^{min}+\epsilon_i)^2\}+\kappa_i(1-\kappa_i)\tilde{S}_{i,t}\epsilon_i+0.5(1-\kappa_i)^2\epsilon_i^2]$ is a constant for the given system parameters $\{\delta_i,\epsilon_i,\kappa_i, S_i^{min},S_i^{max}\}$. $\hfill\blacksquare$

\textbf{Remark 2.4}: For the traditional Lyapunov optimization method \cite{zhong2019online,qin2015online}, optimal operating variable of ESS can only take either the boundary value or zero (e.g., $w_i^{min}$, 0, and $w_i^{max}$), since the optimization problem normally becomes a linear programming with regard to the operating variable. Zero value means that ESS will not take any actions. In order to automatically satisfy constraints \eqref{apen_f11}-\eqref{apen_f12} under the given optimal operating variable obtained by the traditional Lyapunov optimization method, the shifting variable $\epsilon_i$ and weight parameter $\delta_i$ should be well designed beforehand. Different from the traditional method, the square term  $w_{i,t}^2$ is reserved as a non-constant term and makes the optimization problem as a quadratic programming. Meanwhile, the shifting variable $\epsilon_i$ and weight parameter $\delta_i$ no longer need to be well designed since constraints \eqref{apen_f11}-\eqref{apen_f12} are explicitly reserved in the proposed modified Lyapunov optimization method. The operating variable of ESS can there-fore have more choices of optimal decision values within the boundary constraints instead of the limited solutions (e.g., $w_i^{min}$, 0, and $w_i^{max}$). $S_{i,t+1}$ is updated based on the current SoC of ESS, $S_{i,t}$, which is denoted as constraint \eqref{apen_f10}. In this way, constraints \eqref{apen_f11}-\eqref{apen_f12} can be always satisfied during the market operation. The effectiveness is further demonstrated in following sections. The enlarged solution space renders more flexibility for operating ESS and better performance for the real-time P2P energy market operation.

To stabilize the virtual queue as well as simultaneously minimize the objective of the optimization problem, P1 is reformulated into the drift-plus-cost problem \textbf{P2} by considering the non-constant upper bound of the per-time slot Lyapunov drift term obtained from \textbf{Lemma 2.1}:
\begin{subequations}
	\begin{align}
		&\textbf{P2:} \quad \mathop{min}_{\Phi_t} \quad f_t(\mathbf{\Phi}_t) +\sum_{i \in \mathcal{N}_p} \{\delta_i \cdot[\kappa_i\tilde{S}_{i,t}w_{i,t}+ \epsilon_i(1-\kappa_i)w_{i,t}+\frac{w_{i,t}^2}{2}]\} \\
		&\quad\quad\quad s.t. \quad \eqref{apen_f4}-\eqref{apen_f9}, \eqref{apen_f11}-\eqref{apen_f19}.
	\end{align}
\end{subequations}
\textbf{P2} is reformulated from \textbf{P1}, and the time-coupling constraint \eqref{apen_f10} is removed in \textbf{P2}. Thus, \textbf{P2} denotes an online optimization problem to minimize the composite objective at each time slot. \textbf{P2} can be solved based on the previous and current realizations of the system states. The relation between \textbf{P1} and \textbf{P2} and the near-optimal performance guarantee of \textbf{P2} are further given. Before that, we intend to introduce the proposed online distributed algorithm for solving \textbf{P2}, since the distributed algorithm introduces no additional performance gap.

\section{Online Distributed Algorithm for Real-Time P2P Energy Trading}
The reformulated problem \textbf{P2} can be solved in a centralized manner at each time slot by requiring the whole information from all the agents, which inevitably infringes on their privacy. This section proposes a consensus ADMM to tackle the spatial constraints and to solve the problem \textbf{P2} in a distributed manner. Further, the closed-form solutions for the decomposed sub-problems are derived. In this way, private data can be preserved such as the parameters of prosumers’ utility functions and local constraints, and only partial information is shared. 

\subsection{Consensus ADMM Based Distributed Solution}
The spatial physical network constrains are reformulated to constraints \eqref{apen_f4}-\eqref{apen_f9}, and can be handled in a distributed client-server manner by transmitting partial information with the utility company. However, it can be observed that spatial constraint \eqref{apen_f17} makes \textbf{P2} also hard to be solved in a distributed manner. To resolve this issue, a set of auxiliary variables $u_{i,j,t}$ is introduced for each prosumer $i$, which is a duplicate of the original variables $e_{i,j,t}$, $\forall j \in \mathcal{N}_i$ . Let $u_{i,j,t}=e_{i,j,t}$, and then the spatial constraint \eqref{apen_f17} is replaced by the following two relations:
\begin{subequations}
	\begin{align}
		&u_{i,j,t}=e_{i,j,t}, j \in \mathcal{N}_i \label{apen_f21}\\
		&u_{i,j,t}+u_{j,i,t}=0, j \in \mathcal{N}_i. \label{apen_f22}
	\end{align}
\end{subequations}
Accordingly, \textbf{P2} is further reformulated into \textbf{P3}:
\begin{subequations}
	\begin{align}
		&\textbf{P3:} \quad \mathop{min}_{\Phi_t} \quad f_t(\mathbf{\Phi}_t) +\sum_{i \in \mathcal{N}_p} \{\delta_i\cdot [\kappa_i\tilde{S}_{i,t}w_{i,t}+ \epsilon_i(1-\kappa_i)w_{i,t}+\frac{w_{i,t}^2}{2}]\} \\
		& s.t. \quad \eqref{apen_f4}-\eqref{apen_f9}, \eqref{apen_f11}-\eqref{apen_f16}, \eqref{apen_f18}, \eqref{apen_f19},\eqref{apen_f21}, \eqref{apen_f22}.
	\end{align}
\end{subequations}

Constraints \eqref{apen_f7}-\eqref{apen_f9}, \eqref{apen_f11}-\eqref{apen_f16}, \eqref{apen_f18}, and \eqref{apen_f19}, refer to the local constraints. The partial augmented Lagrangian function in respect of constraint \eqref{apen_f21} is written as:
\begin{align}
	&\mathcal{L} (\mathbf{\Phi}_t, \mathbf{u}_t, \mathbf{h}_t)= f_t(\mathbf{\Phi}_t) +\sum_{i \in \mathcal{N}_p} \{\delta_i\cdot [\kappa_i\tilde{S}_{i,t}w_{i,t}+ \epsilon_i(1-\kappa_i)w_{i,t}+\frac{w_{i,t}^2}{2}] \notag\\ 
	&\quad\quad\quad\quad\quad +\sum_{j \in \mathcal{N}_i}\{h_{i,j,t}(u_{i,j,t}-e_{i,j,t})+\frac{\eth}{2}(u_{i,j,t}-e_{i,j,t})^2\}\},
\end{align}
where $\mathbf{u}_t=\{u_{1,1,t,...,u_{1,j,t},...,u_{i,1,t},...,u{i,j,t}}\}$ and $\mathbf{h}=\{h_{1,1,t},...,h_{1,j,t},...,$ $h_{i,1,t},...,h_{i,j,t}\}$. $\eta$ is the penalty factor. $h_{i,j,t}$ is the shadow price for the traded energy $e_{i,j,t}$ between prosumers $i$ and $j$ \cite{le2020peer}. The introduction of the quadratic penalty term $\eta/2\cdot(u_{i,j,t}-e_{i,j,t})^2$ helps to bring robustness to the distributed method such as the consensus ADMM and improve its performance for iteratively updating the primal variables. Large value of $\eta$ tends to produce small primal residuals. Conversely, small value of  $\eta$ tends to reduce the dual residuals and may result in larger primal residuals \cite{boyd2011distributed}. When the distributed method converges, this term tends to be zero.

For each prosumer, \textbf{P3} is decomposed into three sub-problems \textbf{Sp1}, \textbf{Sp2}, and \textbf{Sp3}. Prosumer $i$ should iteratively solve the three optimization problems at each time slot. Each agent is assumed to rationally minimize its own cost function, and meanwhile follow the communication protocols by sharing information with other agents. To begin with, each prosumer $i$ solves the following sub-problem Sp1 to update the primal variables $\mathbf{\Phi}_{i,j,t}^{k+1}=\{e_{i,j,t}^{k+1},p_{i,t}^{k+1},w_{i,t}^{k+1}\}$  for the given auxiliary variable $u_{i,j,t}^k$ and shadow price  $h_{i,j,t}^k$ from the last iteration: 
\begin{subequations}
	\begin{align}
		&\textbf{Sp1:} \quad \mathbf{\Phi}_{i,j,t}^{k+1}=\mathop{argmin}_{\Phi_{i,j,t}} \{\mathcal{L}(\mathbf{\Phi}_t,u_{i,j,t}^k,h_{i,j,t}^k)\} \\
		&\quad\quad\quad s.t. \quad \eqref{apen_f4}-\eqref{apen_f9}, \eqref{apen_f11}-\eqref{apen_f16}, \eqref{apen_f18}, \eqref{apen_f19}.
	\end{align}
\end{subequations}
Then each prosumer $i$ solves the following sub-problem \textbf{Sp2} to update the auxiliary variable   for the newly updated primal variable $e_{i,j,t}^{k+1}$ and the given shadow price $h_{i,j,t}^k$:
\begin{subequations}
	\begin{align}
		&\textbf{Sp2:} \quad u_{i,j,t}^{k+1}=\mathop{argmin}_{u_{i,j,t}} \{ h_{i,j,t}^k(u_{i,j,t}-e_{i,j,t}^{k+1})+\frac{\eta}{2}(u_{i,j,t}-e_{i,j,t}^{k+1})^2 \notag\\
		&\quad\quad\quad\quad\quad\quad +h_{j,i,t}^k(u_{j,i,t}-e_{j,i,t}^{k+1})+\frac{\eta}{2}(u_{j,i,t}-e_{j,i,t}^{k+1})^2\} \\
		&\quad\quad\quad\quad\quad\quad s.t. \quad \eqref{apen_f22}.
	\end{align}
\end{subequations}
Each prosumer $i$ finally solves the following sub-problem \textbf{Sp3} to update the shadow prices $h_{i,j,t}^{k+1}$ for the newly updated primal variable $e_{i,j,t}^{k+1}$ and newly updated auxiliary variable $u_{i,j,t}^{k+1}$:
\begin{align}
	&\textbf{Sp3:} \quad h_{i,j,t}^{k+1}=h_{i,j,t}^{k}+\eta(u_{i,j,t}^{k+1}-e_{i,j,t}^{k+1}). 
\end{align}
Specifically, to stabilize the shadow price and accelerate convergence of the online distributed algorithm, the shadow price is updated in a consensus manner:
\begin{align}
	&h_{i,j,t}^{k+1}=h_{j,i,t}^{k+1}=\frac{1}{2}(h_{i,j,t}^{k+1}+h_{j,i,t}^{k+1}). \label{2-0}
\end{align}
\subsection{Closed-Form Solutions for the Sub-problems}
Closed-form solutions for the first two sub-problems are derived in this part, which can help to improve computational efficiency. The augmented Lagrangian function for all the prosumers in respect of the constraints \eqref{apen_f7}-\eqref{apen_f9}, and \eqref{apen_f12}-\eqref{apen_f14} can be formulated as follows:
\begin{subequations}
\begin{align}
&\mathcal{L}(\mathbf{\Phi}_t,\mathbf{u}_t,\mathbf{h}_t)= \sum_{i \in \mathcal{N}_p} \{(\alpha_i\sum_{j \in \mathcal{N}_i}e_{i,j,t}^2+\beta_i\sum_{j \in \mathcal{N}_i}e_{i,j,t})+\gamma_i(p_{i,t}-g_{i,t}+d_{i,t})^2 \notag\\
	&+\delta_i\cdot[\kappa_i\tilde{S}_{i,t}w_{i,t}+\epsilon_i(1-\kappa_i)w_{i,t}+\frac{w_{i,t}^2}{2}]+|\xi_iw_{i,t}|-\lambda_t^s\cdot p_{i,t}^s+\lambda_t^b\cdot p_{i,t}^b \\
	&+\sum_{j \in \mathcal{N}_i}\{h_{i,j,t}(u_{i,j,t}-e_{i,j,t})+\frac{1}{2}(u_{i,j,t}-e_{i,j,t})^2\}\} \\
	&+\sum_{i \in \mathcal{N}_p} \{\bar{\rho}_{i,p,t}(g_{i,t}-p_{i,t}-d_{i,t}^{max})-\underline{\rho}_{i,p,t}(g_{i,t}-p_{i,t}-d_{i,t}^{min})\} \\
	&+\sum_{i \in \mathcal{N}_p} \{\bar{\rho}_{i,v,t}(v_{i,t}-v_{i}^{max})-\underline{\rho}_{i,v,t}(v_{i,t}-v_{i}^{min})\} \\
	&+\sum_{i \in \mathcal{N}_p} \{\bar{\rho}_{i,P,t}(P_{i,t}-P_{i}^{max})-\underline{\rho}_{i,P,t}(P_{i,t}-P_{i}^{min})\} \\
	&+\sum_{i \in \mathcal{N}_p} \{\bar{\rho}_{i,Q,t}(Q_{i,t}-Q_{i}^{max})-\underline{\rho}_{i,Q,t}(Q_{i,t}-Q_{i}^{min})\} \\
	&+\sum_{i \in \mathcal{N}_p} \{\bar{\rho}_{i,w,t}(w_{i,t}-w_{i}^{max})-\underline{\rho}_{i,w,t}(w_{i,t}-w_{i}^{min})\}.
\end{align}
\end{subequations}
Since the Lagrangian functions for the buyer and seller are distinct due to constraints \eqref{apen_f18} and \eqref{apen_f19}, their closed-form solutions take different forms. To deal with these two constraints, an additional multiplier $\mu_{i,t}$ is introduced. The fully augmented Lagrangian functions for each buyer and seller take the following forms:
\begin{subequations}
	\begin{align}
		&\mathcal{L}_i^p(\mathbf{\Phi}_t,\mathbf{u}_t,\mathbf{h}_t)=\mathcal{L}_i(\mathbf{\Phi}_t,\mathbf{u}_t,\mathbf{h}_t)-\mu_{i,t}(\sum_{j \in \mathcal{N}_s}e_{i,j,t}+w_{i,t}-p_{i,t}) \\
		&\mathcal{L}_i^p(\mathbf{\Phi}_t,\mathbf{u}_t,\mathbf{h}_t)=\mathcal{L}_i(\mathbf{\Phi}_t,\mathbf{u}_t,\mathbf{h}_t)-\mu_{i,t}(p_{i,t}-\sum_{j \in \mathcal{N}_s}e_{i,j,t}-w_{i,t}),
	\end{align}
\end{subequations}
where $\mathcal{L}_i(\mathbf{\Phi}_t,\mathbf{u}_t,\mathbf{h}_t)$ is the splitting augmented Lagrangian function only relevant to prosumer $i$.
By applying the Karush-Kuhn-Tucker (KKT) conditions, the closed-form solutions of the primal problem \textbf{Sp1} are obtained by letting $\triangledown_{e_{i,j,t}}\mathcal{L}_i^p(\mathbf{\Phi}_t,\mathbf{u}_t,\mathbf{h}_t)=0$. Then the closed-form solutions for each buyer and seller are obtained in respect of the optimal traded energy $e_{i,j,t}^{k+1}$:
\begin{subequations}
	\begin{align}
		&e_{i,j,t}^{k+1}=\frac{-\beta-\lambda_t^b+\mu_{i,t}^k+h_{i,t}^k+\eta u_{i,j,t}^k}{2\alpha_i+\eta}, i \in \mathcal{N}_b, j \in \mathcal{N}_i \label{2-1}\\
		&e_{i,j,t}^{k+1}=\frac{-\beta-\lambda_t^s-\mu_{i,t}^k+h_{i,t}^k+\eta u_{i,j,t}^k}{2\alpha_i+\eta}, i \in \mathcal{N}_s, j \in \mathcal{N}_i. \label{2-2}
	\end{align}
\end{subequations}
Similarly, let $\triangledown_{w_{i,t}}\mathcal{L}_i^p(\mathbf{\Phi}_t,\mathbf{u}_t,\mathbf{h}_t)=0$. The closed-form solutions of the optimal quantity of charging/discharging energy of ESS for each buyer and seller are obtained when taking the boundary constraint \eqref{apen_f11} into consideration, which are denoted as:
\begin{subequations}
	\begin{align}
		&w_{i,t}^{k+1}=\begin{cases}\resizebox{0.8\hsize}{!}{$ min\{[-\frac{\xi_i+(\kappa_iS_{i,t}+\epsilon_i)\cdot \delta_i+\lambda_t^b-\mu_{i,t}^k+\bar{\rho}_{i,w,t}^k-\underline{\rho}_{i,w,t}^k}{\delta_i}]^+,S_i^{max}-\kappa_i S_{i,t}\}, $ if $ w_{i,t}^{k+1} \ge 0 $}, i \in \mathcal{N}_b \\
			\resizebox{0.8\hsize}{!}{$ max\{-[\frac{-\xi_i+(\kappa_iS_{i,t}+\epsilon_i)\cdot \delta_i+\lambda_t^b-\mu_{i,t}^k+\bar{\rho}_{i,w,t}^k-\underline{\rho}_{i,w,t}^k}{\delta_i}]^+,S_i^{min}-\kappa_i S_{i,t}\}, $ if $ w_{i,t}^{k+1} < 0 $}, i \in \mathcal{N}_b
		\end{cases} \label{2-3}\\
		&w_{i,t}^{k+1}=\begin{cases}\resizebox{0.8\hsize}{!}{$ min\{[-\frac{\xi_i+(\kappa_iS_{i,t}+\epsilon_i)\cdot \delta_i+\lambda_t^s-\mu_{i,t}^k+\bar{\rho}_{i,w,t}^k-\underline{\rho}_{i,w,t}^k}{\delta_i}]^+,S_i^{max}-\kappa_i S_{i,t}\}, $ if $ w_{i,t}^{k+1} \ge 0 $}, i \in \mathcal{N}_s \\
			\resizebox{0.8\hsize}{!}{$ max\{-[\frac{-\xi_i+(\kappa_iS_{i,t}+\epsilon_i)\cdot \delta_i+\lambda_t^s-\mu_{i,t}^k+\bar{\rho}_{i,w,t}^k-\underline{\rho}_{i,w,t}^k}{\delta_i}]^+,S_i^{min}-\kappa_i S_{i,t}\}, $ if $ w_{i,t}^{k+1} < 0 $}, i \in \mathcal{N}_s.
		\end{cases} \label{2-4}
	\end{align}
\end{subequations}
As is observed, the closed-form solutions for the decision variable of ESS, $w_{i,t}$, can take the feasible values within its constraints other than limited solutions from the traditional Lyapunov optimization method, and the effectiveness is further demonstrated from simulation results of the case D and case E.

By letting $\triangledown_{p_{i,t}}\mathcal{L}_i^p(\mathbf{\Phi}_t,\mathbf{u}_t,\mathbf{h}_t)=0$, the closed-form solutions of the optimal active power injection $p_{i,t}^{k+1}$ for each buyer and seller are obtained:
\begin{subequations}
	\begin{align}
		&p_{i,t}^{k+1}=\frac{2\gamma_i(g_{i,t}-d_{i,t})+\lambda_t^b-\mu_{i,t}^k+\bar{\rho}_{i,p,t}^k-\underline{\rho}_{i,p,t}^k-\mathcal{R}_{i,t}^k}{2\gamma_i}, i \in \mathcal{N}_b \label{2-5}\\
		&p_{i,t}^{k+1}=\frac{2\gamma_i(g_{i,t}-d_{i,t})+\lambda_t^s+\mu_{i,t}^k+\bar{\rho}_{i,p,t}^k-\underline{\rho}_{i,p,t}^k-\mathcal{R}_{i,t}^k}{2\gamma_i}, i \in \mathcal{N}_s, \label{2-6}
	\end{align}
\end{subequations}
where the term $\mathcal{R}_{i,t}^k$ is denoted as follows:
\begin{subequations}
	\begin{align}
		&\mathcal{R}_{i,t}^k=\sum_{r \in \mathcal{N}_p}\{\mathcal{Z}_{r,i}(\bar{\rho}_{r,v,t}^k-\underline{\rho}_{r,v,t}^k)+\mathcal{B}_{r,i}(\bar{\rho}_{r,P,t}-\underline{\rho}_{r,P,t})+\mathcal{X}_i\mathcal{B}_{r,i}(\bar{\rho}_{r,Q,t}-\underline{\rho}_{r,Q,t})\} \label{2-7}\\
		&\mathcal{Z}=2\mathcal{H}^{-1}D_r\mathcal{H}^{-T}+2\mathcal{H}^{-1}D_x\mathcal{H}^{-T}\mathcal{X}_i, \mathcal{B}=\mathcal{H}^{-T}. \label{2-8}
	\end{align} 
\end{subequations}
The closed-form optimal solution for the sub-problem \textbf{Sp2} is obtained by applying the KKT conditions to \textbf{Sp2}:  
\begin{align}
	&u_{i,j,t}^{k+1}=-u_{j,i,t}^{k+1}=\frac{\eta(e_{i,j,t}^{k+1}-e_{j,i,t}^{k+1})-(h_{i,j,t}^k-h_{j,i,t}^k)}{2\eta}. \label{2-9}
\end{align}
The dual multipliers $\bar{\rho}_{i,p,t}^k$ and $\underline{\rho}_{i,p,t}^k$ are updated ascendingly by introducing a constant parameter $\tau_p$: 
\begin{subequations}
	\begin{align}
		&\bar{\rho}_{i,p,t}^k=[\bar{\rho}_{i,p,t}^k-\tau_p(p_{i,t}^{k+1}-g_{i,t}+d_{i,t}^{max})]^+ \label{2-101}\\
		&\underline{\rho}_{i,p,t}^k=[\underline{\rho}_{i,p,t}^k+\tau_p(p_{i,t}^{k+1}-g_{i,t}+d_{i,t}^{min})]^+. \label{2-10}
	\end{align}
\end{subequations}
Similarly, the dual multiplier $u_{i,t}^k$ are updated ascendingly by introducing a constant parameter $\tau_u$: 
\begin{subequations}
	\begin{align}
		&\mu_{i,t}^{k+1}=[\mu_{i,t}^{k}-\tau_u(\sum_{i \in \mathcal{N}_s} e_{i,j,t}^{k+1}-w_{i,t}^{k+1}-p_{i,t}^{k+1})]^+, i \in \mathcal{N}_b \label{2-11}\\
		&\mu_{i,t}^{k+1}=[\mu_{i,t}^{k}-\tau_u(p_{i,t}^{k+1}-\sum_{i \in \mathcal{N}_b} e_{i,j,t}^{k+1}-w_{i,t}^{k+1})]^+, i \in \mathcal{N}_s. \label{2-12}
	\end{align}
\end{subequations}
For the other dual multipliers, they are updated by introducing a vector of the constant step parameters $\tau_{\mathbf{f}}$ :      
\begin{subequations}
	\begin{align}
		&\bar{\boldsymbol{\rho}}^{k+1}=[\bar{\boldsymbol{\rho}}^{k}+\tau_{\mathbf{f}}(\mathbf{f}^{k+1}-\mathbf{f}^{max})]^+ \label{2-13}\\
		&\underline{\boldsymbol{\rho}}^{k+1}=[\underline{\boldsymbol{\rho}}^{k}-\tau_{\mathbf{f}}(\mathbf{f}^{k+1}-\mathbf{f}^{min})]^+, \label{2-14}
	\end{align}
\end{subequations}
where  $\bar{\boldsymbol{\rho}}=\{\bar{\rho}_{i,v,t},\bar{\rho}_{i,P,t},\bar{\rho}_{i,Q,t},\bar{\rho}_{i,w,t}\}$, $\underline{\boldsymbol{\rho}}=\{\underline{\rho}_{i,v,t},\underline{\rho}_{i,P,t},\underline{\rho}_{i,Q,t},\underline{\rho}_{i,w,t}\}$, and $ \boldsymbol{f}=\{v_{i,t},P_{i,t},Q_{i,t},w_{i,t}\}$. 

\subsection{Convergence Criterion for the Online Distributed Algorithm}
The convergence criterion of the proposed online distributed Algorithm \ref{apen_a1} as detailly illustrated in next part takes the primal and auxiliary variables into account, and the composite residual term takes on the following form:
\begin{align}
	r_{t}^{k+1}=\sqrt{\sum_{i \in \mathcal{N}_p}\sum_{j \in \mathcal{N}_i}\{(e_{i,j,t}^{k+1}+e_{j,i,t}^{k+1})^2+(u_{i,j,t}^{k+1}-e_{i,j,t}^{k+1})^2+(u_{i,j,t}^{k+1}-u_{i,j,t}^k)^2\}}. \label{2-15}
\end{align}
The stopping condition for the online distributed algorithm is set to be $r_{t}^{k+1} \le 10^{-3}$. The proposed algorithm employs the consensus ADMM, which guarantees the $o(1/k)$ convergence rate. The proof can be found in \cite{boyd2011distributed}, and detailed analysis is omitted here.

\subsection{Proposed Online Distributed Algorithm}
\begin{figure}[!hp]
	\centering
	\includegraphics[width=1.0\linewidth]{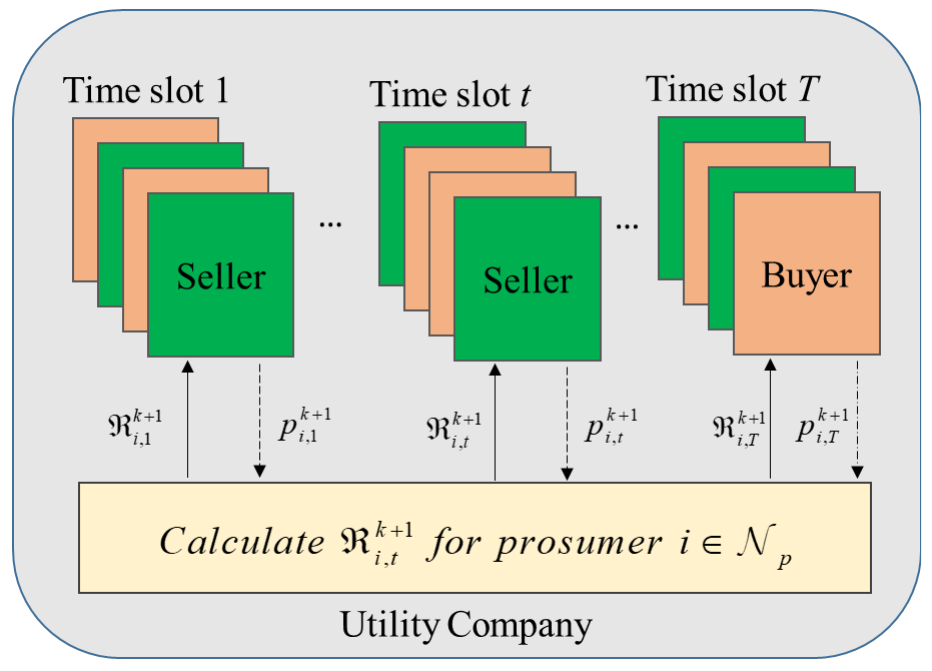}
	\caption{Information exchange between the prosumers and utility company over the operation period.}
	\label{apen_2}
\end{figure}

Since utility company is responsible for the operation of local distribution power network, the utility company is additionally introduced as a network manager to compute the term $R_{i,t}^{k+1}$ related to the network constraints for each prosumer $i$ over the operation period, as shown in Figure \ref{apen_2}. As denoted in the equations \eqref{2-7} and \eqref{2-8}, $R_{i,t}^{k+1}$ is composed of the shadow prices, such as $\bar{\rho}_{i,v,t},\bar{\rho}_{i,P,t},\bar{\rho}_{i,Q,t},\underline{\rho}_{i,v,t},\underline{\rho}_{i,P,t}$, and $\underline{\rho}_{i,Q,t}$, which are the dual variables about the bus voltage, active power, and reactive power over the lines. The term $R_{i,t}^{k+1}$ is therefore can be considered as the price that accounts for the marginal costs for the voltage support and the line congestion. 

In addition, each prosumer needs to transmit their active power injection $p_{i,t}^{k+1}$ to the utility company and receive the network-related term $R_{i,t}^{k+1}$ at each iteration of every time slot. Under this setting, some private information, such as the parameters of the cost functions and local constraints, are preserved. After receive all the active power injection $p_{i,t}^{k+1}$, the utility company can calculate the network-related term $R_{i,t}^{k+1}$, and send this term to each prosumer. This term is performed as the feedback during each iteration of every time slot. Meanwhile, it also functions as a penalty term that automatically regulates the active power injection $p_{i,t}^{k+1}$ and other decision variables such as the traded power and the amount of charging energy for ESS in order to meet the network constraints. The proposed Algorithm \ref{apen_a1} ensures that the real-time P2P energy market operates in an online and distributed manner. At a given time slot, only the partial information of each prosumer is required to share with other prosumers such as $h_{i,j,t}^{k+1}$ and $e_{i,j,t}^{k+1}$ at each iteration, as illustrated in the Algorithm \ref{apen_a1}. The pro-posed online distributed method employs a mixed distributed architecture with peer-to-peer and client-server communications, since each prosumer needs to exchange information not only with other prosumers but with a server node, i.e., the utility company. For the convergence criterion, an additional maximum allowed iteration, $k^{max}$, is introduced for the verification of the convergence.

\begin{algorithm}[H]
	\caption{Consensus ADMM-based Online Distributed Algorithm. \cite{boyd2011distributed}}\label{apen_a1}
	\begin{algorithmic}
		\STATE 
		\STATE {\textbf{Initialization:}}
		\STATE \hspace{0.1cm}$ \text{Initialize } e_{i,j,t}^1, w_{i,t}^1, p_{i,t}^1, h_{i,j,t}^1, u_{i,j,t}^1, \mathcal{R}_{i,t}^1, \boldsymbol{\bar{\rho}}^1, \underline{\boldsymbol{\rho}}^1, k,t, \mathcal{T}, k_{max}$ 
		\STATE {\textbf{Algorithm for each prosumer} $i \in \mathcal{N}_p$}
		\STATE \hspace{0.1cm}$ \textbf{for } t \le \mathcal{T} \textbf{ do}$ 
		\STATE \hspace{0.3cm}$ \textbf{while } k \le k^{max} \textbf{ do}$ 
		\STATE \hspace{0.5cm} $\text{Update } e_{i,j,t}^{k+1} \text{ by \eqref{2-1} or \eqref{2-2}},  w_{i,t}^{k+1} \text{ by \eqref{2-3} or \eqref{2-4}}, $
		\STATE \hspace{0.5cm}$ \qquad p_{i,t}^{k+1}  \text{ by \eqref{2-5} or \eqref{2-6}, and  } \mu_{i,t}^{k+1} \text{  by \eqref{2-11} or \eqref{2-12}};$ 
		\STATE \hspace{0.5cm} $\text{Update } u_{i,j,t}^{k+1} \text{ by \eqref{2-9}, } h_{i,j,t}^{k+1} \text{ by \eqref{2-0}, } \bar{\rho}_{i,p,t}^{k+1} \text{ by \eqref{2-101}, }  $
		\STATE \hspace{0.5cm}$\qquad \underline{\rho}_{i,p,t}^{k+1} \text{ by \eqref{2-10}, } \bar{\rho}_{i,w,t}^{k+1} \text{ by \eqref{2-13}, and } \underline{\rho}_{i,w,t}^{k+1} \text{ by \eqref{2-14}};$
		\STATE \hspace{0.5cm} $\text{Exchange } e_{i,j,t}^{k+1} \text{,} e_{j,i,t}^{k+1} \text{ with prosumer }  j \in \mathcal{N}_i \text{ at iteration } k;$
		\STATE \hspace{0.5cm} $\text{Exchange } h_{i,j,t}^{k+1} \text{,} h_{j,i,t}^{k+1} \text{  with prosumer } j \in \mathcal{N}_i \text{ at iteration } k;$
		\STATE \hspace{0.5cm} $\text{Transmit }  p_{i,t}^{k+1}  \text{ to utility company at iteration } k  \text{ of slot } t;$\\
		\STATE \hspace{0.5cm} $\text{Receive } \mathcal{R}_{i,t}^{k+1} \text{ from utility company at iteration } k \text{ of slot } t;$	
		\STATE \hspace{0.5cm} $\text{Set } k \leftarrow k+1 $	  
		\STATE \hspace{0.3cm}\textbf{end}	
		\STATE \hspace{0.3cm}$\text{Set } S_{i,t+1} \leftarrow \kappa_i S_{i,t}+w_{i,t}, \text{ and } t \leftarrow t+1$	 
		\STATE \hspace{0.1cm}\textbf{end}   
		\STATE {\textbf{Algorithm for utility company}}
		\STATE \hspace{0.1cm}$ \textbf{for } t \le \mathcal{T} \textbf{ do}$ 
		\STATE \hspace{0.3cm}$ \textbf{while } k \le k^{max} \textbf{ do}$ 
		\STATE \hspace{0.5cm} $\text{Receive } p_{i,t}^{k+1} \text{ from prosumer } i \text{ at iteration } k  \text{ of slot } t;$ 
		\STATE \hspace{0.5cm} $\text{Update } \bar{\rho}_{i,v,t},\bar{\rho}_{i,P,t},\bar{\rho}_{i,Q,t},\underline{\rho}_{i,v,t},\underline{\rho}_{i,P,t}, \underline{\rho}_{i,Q,t}, \forall i \in \mathcal{N}_p;$
		\STATE \hspace{0.5cm} $\text{Calculate } \mathcal{R}_{i,t}^{k+1}  \text{ by \eqref{2-7} and \eqref{2-8} for prosumer } i \in \mathcal{N}_p;$
		\STATE \hspace{0.5cm} $\text{Transmit }  \mathcal{R}_{i,t}^{k+1}  \text{ to prosumer } i \in \mathcal{N}_p  \text{ at iteration } k  \text{ of slot } t;$ 
		\STATE \hspace{0.5cm} $k \leftarrow k+1$	
		\STATE \hspace{0.3cm}\textbf{end}	
		\STATE \hspace{0.3cm}$t \leftarrow t+1$ 
		\STATE \hspace{0.1cm} \textbf{end}
	\end{algorithmic}
\end{algorithm}

\section{Performance Guarantee for Online Distributed Algorithm}
The original spatial-temporally constrained stochastic optimization problem P1 is reformulated to the online optimization problem \textbf{P2} by relaxing the time-coupling constraints, and the performance gap remains. However, the reformulation from \textbf{P2} to \textbf{P3} by the consensus ADMM is nearly exact and zero performance gap remains, which means that \textbf{P2} is equivalent to \textbf{P3} in respect of the performance. Thus, the performance guarantee for the online distributed algorithm is simplified to focus on the reformulation step from \textbf{P1} to \textbf{P2}. In this section, part A provides the derivation of the theoretical near-optimal performance guarantee for the proposed online distributed algorithm.

\subsection{Performance Guarantee for the Proposed Online Distributed Algorithm }
The time-averaged values for the SoC and the charging/discharging energy of ESS for prosumer $i$ over the online operation period are defined as follows:
\begin{subequations}
	\begin{align}
		& \bar{w}_i \mathop{=}^{\triangle} \mathop{lim}_{\mathcal{T} \rightarrow \infty} \frac{1}{\mathcal{T}} \mathbb{E}[\sum_{t=1}^{t=\mathcal{T}}w_{i,t}]=\mathop{lim}_{\mathcal{T} \rightarrow \infty}\frac{1}{\mathcal{T}}  \mathbb{E}[\sum_{t=1}^{t=\mathcal{T}}S_{i,t+1}-\kappa_i S_{i,t}] \\
		& \bar{S}_i\mathop{=}^{\triangle}\mathop{lim}_{\mathcal{T} \rightarrow \infty} \frac{1}{\mathcal{T}}[\sum_{t=1}^{t=\mathcal{T}}S_{i,t}].
	\end{align}
\end{subequations}
Since the SoC of ESS, $S_{i,t}$, is bounded by $[S_i^{min},S_i^{max}]$, the time-averaged SoC, $\bar{S}_i$, takes the same bounds. In this way, the time-averaged charging/discharging energy of ESS is bounded by:
\begin{align}
	(1-\kappa_i)S_i^{min} \le \bar{w}_i \le (1-\kappa_i)S_i^{max}.
\end{align}
To obtain the performance bound for the online optimization problem \textbf{P2}, the original problem \textbf{P1} is additionally reformulated to \textbf{P4} as follows:
\begin{subequations}
	\begin{align}
		& \textbf{P4}: \quad \mathop{min}_{\mathbf{\Phi}_t} \mathop{lim}_{\mathcal{T} \rightarrow \infty} \frac{1}{\mathcal{T}} \sum_{t=1}^{t=\mathcal{T}} \{\mathbb{E}[f_t(\mathbf{\Phi}_t)]\} \\
		& \qquad s.t. \quad \eqref{apen_f4}-\eqref{apen_f9}, \eqref{apen_f11}-\eqref{apen_f19}  \\
		& \qquad\qquad  (1-\kappa_i)S_i^{min} \le \bar{w}_i \le (1-\kappa_i)S_i^{max}.
	\end{align}
\end{subequations}
Let $\Psi_{P1}^*(\mathbf{\Phi})$ and $\Psi_{P4}^*(\mathbf{\Phi})$ be the optimal objective values of \textbf{P1} and \textbf{P4}, respectively, where $\mathbf{\Phi}=\{\mathbf{\Phi_1},\mathbf{\Phi_t},...,\mathbf{\Phi_T}\}$. Note that \textbf{P4} is an approximate problem of \textbf{P1}, and it can be verified that  $\Psi_{P4}^*(\mathbf{\Phi}) \le \Psi_{P1}^*(\mathbf{\Phi})$ \cite{zhou2019hierarchical}. Meanwhile, the optimal solution of \textbf{P4} may not be feasible to original optimization problem \textbf{P1}. However, it can help bridge the performance gap from \textbf{P1} to \textbf{P2}.

\textbf{Remark 2.5}: In this paper, we suppose that the system state, denoted as  $\mathbf{R}_t=\{g_{1,t},d_{1,t},...,g_{N,t},d_{N,t},\lambda_t^s,\lambda_t^b\}$, is independent and identically distributed (i.i.d.) over the online operation period. In fact, this assumption can be extended to none-i.i.d. situations, such that $\mathcal{R}_t$ follows the finite ergodic Markov chain process \cite{neely2022stochastic}.

\textbf{Lemma 2.2}: Under the i.i.d. assumption for the system states, there exists the stationary solution $\mathbf{\Phi}_t^{\dagger}$ for \textbf{P2} that satisfies the following relations:
\begin{subequations}
	\begin{align}
		& (1-\kappa_i)S_i^{min} \le \mathbb{E}[w_{i,t}^{\dagger}] \le (1-\kappa_i)S_i^{max} \label{apen_f26}\\
		& 0 \le \mathbb{D}[w_{i,t}^{\dagger}] \le (w_i^{max}-w_i^{min})^2  \label{apen_f27}\\
		& \mathbb{E}[f_t(\mathbf{\Phi}_t)|\mathbf{\Phi}_t^{\dagger}]=\Psi_{P4}^*(\mathbf{\Phi}). \label{apen_f28}
	\end{align}
\end{subequations}
\textit{Proof}: For relations \eqref{apen_f26} and \eqref{apen_f28}, detailed proof can be referred to \cite{tushar2021peer} \cite{qin2014modeling,qin2015online,neely2022stochastic}. For relation \eqref{apen_f27}, it can be obtained as follows: 
The statistic variance for $w_{i,t}^{\dagger}$ can be calculated by $\mathbb{D}[w_{i,t}^{\dagger}]=\sum (w_{i,t}^{\dagger}-\bar{w}_{i,t}^{\dagger})/N^{\dagger}$, where $N^{\dagger}$ is number of samples which is assumed to be large enough. Since $w_i^{min}\le w_{i,t}^{\dagger} \le w_i^{max}$, $\mathbb{D}[w_{i,t}^{\dagger}]$ is bounded by   $0 \le \mathbb{D}[w_{i,t}^{\dagger}] \le (w_i^{max}-w_i^{min})^2$. $\hfill\blacksquare$

\textbf{Theorem 2.1}: If the system state   is i.i.d. over the online operation period, time-averaged system cost of the original problem \textbf{P1} under the optimal online decision sequence $\mathbf{\Phi}{*}=\{\mathbf{\Phi}_1^{*},\mathbf{\Phi}_2^{*},...,\mathbf{\Phi}_T^{*}\}$ generated by \textbf{P2} can be bounded based on \textbf{Lemma 2.2}:
\begin{equation}
	\Psi_{P1}^*(\mathbf{\Phi}) \le \Psi_{P1}(\mathbf{\Phi}^*) \le \Psi_{P1}^*(\mathbf{\Phi})+\Theta,
\end{equation}
where
\begin{subequations}
	\begin{align}	
		\Theta=&\sum_{i \in \mathcal{N}_p} \{\delta_i(\kappa_i-\frac{1}{2}-\frac{1}{2}\kappa_i^2)\cdot max\{(S_i^{min}+\epsilon_i)^2,(S_i^{max}+\epsilon_i)^2\}+ \\
		&\frac{1}{2}\delta_i\cdot(\kappa_i-1)^2\cdot(S_i^{max})^2+\frac{1}{2}\delta_i(1-\kappa_i)^2\epsilon_i^2+\frac{1}{2}\delta_i(w_i^{max}-w_i^{min})^2\}.
	\end{align}
\end{subequations}

\textit{Proof}: Let $\mathbf{\Phi}_t^*$ be the optimal solution of \textbf{P2} at time slot $t$, and then we can derive the following relations:
\begin{subequations}
	\begin{align}	
		&\sum_{i \in \mathcal{N}_p} \{\triangle(\tilde{S}_{i,t})\}+\mathbb{E}[f_t(\mathbf{\Phi}_t)|\tilde{S}_t,\mathbf{\Phi}_t^*] \le \sum_{i \in \mathcal{N}_p}\{M_i\}+\mathbb{E}[\Psi_{P2}^*(\mathbf{\Phi}_t)|\tilde{S}_t] \\
		&\le \sum_{i \in \mathcal{N}_p}\{\delta_i\cdot \mathbb{E}[\frac{1}{2}(\kappa_i^2-1)\cdot max\{(S_i^{max}+\epsilon_i)^2,(S_i^{min}+\epsilon_i)^2\}+ \notag \\
		&\frac{1}{2}(1-\kappa_i)^2\epsilon_i^2|\mathbf{\Phi}_t^{\dagger}]\}+\sum_{i \in \mathcal{N}_p}\{\mathbb{E}[\delta_i \kappa_i(1-\kappa_i)\tilde{S}_{i,t}\epsilon_i]\}+\mathbb{E}[\Psi_{P2}(\mathbf{\Phi}_t)|\mathbf{\Phi}_t^{\dagger}] \\
		&\le \Theta_1+\sum_{i \in \mathcal{N}_p}\{\mathbb{E}[\delta_i\kappa_i(1-\kappa_i)\tilde{S}_{i,t}\epsilon_i+\delta_i\kappa_iw_{i,t}\tilde{S}_{i,t}|\boldsymbol{\Phi}_t^{\dagger}]\} \notag\\
		&+\sum_{i \in \mathcal{N}_p}\{\delta_i\cdot \mathbb{E}[\frac{w_{i,t}^2}{2}+(1-\kappa_i)w_{i,t}\epsilon_i|\boldsymbol{\Phi}_t^{\dagger}]\}+\mathbb{E}[f_t(\boldsymbol{\Phi}_t)|\boldsymbol{\Phi}_t^{\dagger}] \\
		&\le \Theta_1+\sum_{i \in \mathcal{N}_p}\{\mathbb{E}[\delta_i\kappa_i\tilde{S}_{i,t}((1-\kappa_i)\epsilon_i+w_{i,t})|\boldsymbol{\Phi}_t^{\dagger}]\} \notag\\
		&+\sum_{i \in \mathcal{N}_p}\{\delta_i\cdot \mathbb{E}[\frac{w_{i,t}^2}{2}+(1-\kappa_i)w_{i,t}\epsilon_i|\boldsymbol{\Phi}_t^{\dagger}]\}+\mathbb{E}[f_t(\boldsymbol{\Phi}_t)|\boldsymbol{\Phi}_t^{\dagger}] \\
		&\le \Theta_1+\Theta_2+\sum_{i \in \mathcal{N}_p}\{\delta_i\cdot \mathbb{E}[\frac{w_{i,t}^2}{2}+(1-\kappa_i)w_{i,t}\epsilon_i|\boldsymbol{\Phi}_t^{\dagger}]\}+\mathbb{E}[f_t(\boldsymbol{\Phi}_t)|\boldsymbol{\Phi}_t^{\dagger}] \\
		&\le \Theta_1+\Theta_2+\Theta_3+\Theta_4+\mathbb{E}[f_t(\boldsymbol{\Phi}_t)|\boldsymbol{\Phi}_t^{\dagger}] \\
		&\le \Theta+\mathbb{E}[f_t(\boldsymbol{\Phi}_t)|\boldsymbol{\Phi}_t^{\dagger}] \\
		&=\Theta+\Psi_{P4}^*(\boldsymbol{\Phi}_t) \\
		&\le \Theta+\Psi_{P1}^*(\boldsymbol{\Phi}_t),
	\end{align}
\end{subequations}
where $\Theta_1=\sum_{i \in \mathcal{N}_p}\{\delta_i\cdot [\frac{1}{2}(\kappa_i^2-1)\cdot max\{(S_i^{max}+\epsilon_i)^2,(S_i^{min}+\epsilon_i)^2\}+\frac{1}{2}(1-\kappa_i)^2\epsilon_i^2]\}$, $\Theta_2=\sum_{i \in \mathcal{N}_p}\{\delta_i\kappa_i(1-\kappa_i)\cdot max\{(S_i^{min}+\epsilon_i)^2,(S_i^{max}+\epsilon_i)^2\}\}$, $\Theta_3=\sum_{i \in \mathcal{N}_p}\{\frac{1}{2}\delta_i\cdot((1-\kappa_i)^2\cdot (S_i^{max})^2+(w_i^{max}-w_i^{min})^2)\}$, $\Theta_4=0$, and $\Theta=\Theta_1+\Theta_2+\Theta_3+\Theta_4$. The terms $\Theta_2$, $\theta_3$, and $\theta_4$ can be derived as follows:
\begin{subequations}
	\begin{align}
		&\sum_{i \in \mathcal{N}_p} \{\mathbb{E}[\delta_i\kappa_i\tilde{S}_{i,t}((1-\kappa_i)\epsilon_i+w_{i,t})|\boldsymbol{\Phi}_t^{\dagger}]\} \\
		&=\sum_{i \in \mathcal{N}_p}\{\mathbb{E}[\delta_i\kappa_i\tilde{S}_{i,t}(1-\kappa_i)\sqrt{max{\{(S_i^{min}+\epsilon_i)^2,(S_i^{max}+\epsilon_i)^2\}}}|\boldsymbol{\Phi}_t^{\dagger}]\} \\
		&\le \sum_{i \in \mathcal{N}_p}\{\delta_i\kappa_i(1-\kappa_i)\cdot max\{(S_i^{min}+\epsilon_i)^2,(S_i^{max}+\epsilon_i)^2\}\} \\
		&=\Theta_2,
	\end{align}
\end{subequations}		

\begin{subequations}
	\begin{align}
		&\sum_{i \in \mathcal{N}_p} \{\delta_i\cdot \mathbb{E}[\frac{w_{i,t}^2}{2}|\boldsymbol{\Phi}_t^{\dagger}]\} \\
		&=\sum_{i \in \mathcal{N}_p}\{\frac{1}{2}\delta_i\cdot(\mathbb{E}[w_{i,t}|\boldsymbol{\Phi}_t^{\dagger}]^2+\mathbb{D}[w_{i,t}|\boldsymbol{\Phi}_t^{\dagger}])\} \\
		&\le \sum_{i \in \mathcal{N}_p}\{\frac{1}{2}\delta_i\cdot((1-\kappa_i)^2\cdot max\{(S_i^{min})^2,(S_i^{max})^2\}+(w_i^{max}-w_i^{min})^2)\} \\
		&=\sum_{i \in \mathcal{N}_p}\{\frac{1}{2}\delta_i\cdot((1-\kappa_i)^2\cdot (S_i^{max})^2+(w_i^{max}-w_i^{min})^2)\} \\
		&=\Theta_3, \\ 
		
		&\sum_{i \in \mathcal{N}_p} \{\delta_i\cdot \mathbb{E}[(1-\kappa_i)w_{i,t}\epsilon_i|\boldsymbol{\Phi}_t^{\dagger}]\} \\
		&\le \sum_{i \in \mathcal{N}_p}\{\delta_i\cdot(1-\kappa_i)^2\cdot S_{i,\lambda}\cdot \epsilon_i\} \\
		&=\Theta_4,
	\end{align}
\end{subequations}
where $S_i^{min}\le S_{i,\lambda} \le S_i^{max}$. Since $\delta_i$ and $S_{i,\lambda}$ usually take the positive values, while the shifting parameter $\epsilon_i$ takes the negative value \cite{paudel2018peer}. The term   $\Theta_4$ can be further bounded by 0. 
In short, the first derived relation can be simplified as $\sum_{i \in \mathcal{N}_p}\{\triangle(\tilde{S}_{i,t})\}+\mathbb{E}[f_t(\boldsymbol{\Phi}_t)|\tilde{S}_{i,t},\boldsymbol{\Phi}_t^*]\le \Theta+\Psi_{P1}^*(\boldsymbol{\Phi}_t)$, which can be further transformed as $\sum_{i \in \mathcal{N}_p}\{\tilde{S}_{i,t+1}-\tilde{S}_{i,t}\}+\mathbb{E}[f_t(\boldsymbol{\Phi}_t)|\tilde{S}_{i,t},\boldsymbol{\Phi}_t^*]\le \Theta+\Psi_{P1}^*(\boldsymbol{\Phi}_t)$. If we sum it up from 1 to $\mathcal{T}$, then we can obtain $\sum_{i \in \mathcal{N}_p}\{\tilde{S}_{i,t+1}-\tilde{S}_{i,t}\}+\sum_{t=1}^{t=\mathcal{T}}\mathbb{E}[f_t(\boldsymbol{\Phi}_t)|\tilde{S}_{i,t},\boldsymbol{\Phi}_t^*]\le \mathcal{T}\cdot \Theta+\sum_{t=1}^{t=\mathcal{T}}\{\Psi_{P1}^*(\boldsymbol{\Phi}_t)\}$. Divide the above formula by $\mathcal{T}$, and we can finally obtain the following relation:
\begin{align} \resizebox{1.0\hsize}{!}{$
		\frac{1}{\mathcal{T}}\sum_{t=1}^{t=\mathcal{T}}\{\mathbb{E}[f_t(\boldsymbol{\Phi}_t)|\tilde{S}_{i,t},\boldsymbol{\Phi}_t^*]\} \le \frac{1}{\mathcal{T}}\sum_{t=1}^{t=\mathcal{T}}\{\Psi_{P1}^*(\boldsymbol{\Phi}_t)\}+\Theta+\frac{1}{\mathcal{T}}\sum_{i \in \mathcal{N}_p}\{\tilde{S}_{i,t+1}-\tilde{S}_{i,t}\}. $}
\end{align}
which is equivalent to $\Psi_{P1}(\boldsymbol{\Phi}^*)\le \Psi_{P1}^*(\boldsymbol{\Phi})+\Theta$ when  $\mathcal{T}\rightarrow \infty$. For the optimization problem \textbf{P1}, since $\Psi_{P1}^*(\boldsymbol{\Phi})$ should be the minimum objective value for any decision sequence $\boldsymbol{\Phi}$, we can obtain $\Psi_{P1}^*(\boldsymbol{\Phi}) \le \Psi_{P1}(\boldsymbol{\Phi}^*) $. Finally, we can conclude Theorem 2.1 as $\Psi_{P1}^*(\boldsymbol{\Phi}) \le \Psi_{P1}(\boldsymbol{\Phi}^*) \le \Psi_{P1}^*(\boldsymbol{\Phi})+\Theta$. $\hfill\blacksquare$

\begin{figure*}[htbp]
	\centering
	\includegraphics[width=0.8\linewidth]{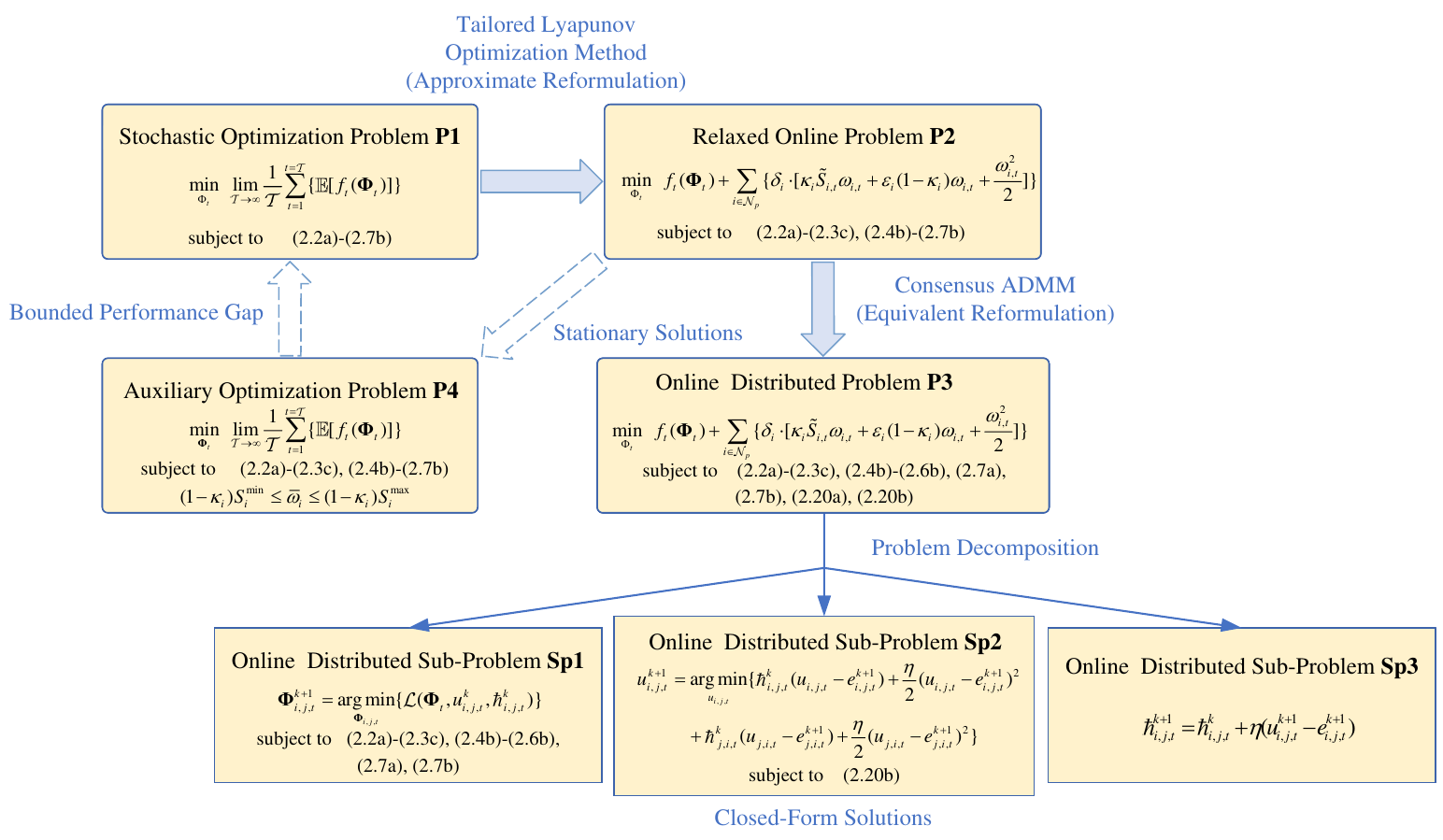}
	\caption{Scheme of the proposed online distributed algorithm.}
	\label{apen_3}
\end{figure*}

The statistic variance of the charging/discharging variable $w_{i,t}^{\dagger}$ is introduced for the kept term $w_{i,t}^2$ from the proposed modified Lyapunov optimization method, as denoted in relation \eqref{apen_f27}, which helps to establish the theoretical performance bound $\Theta$ for \textbf{P2}. By constructing an auxiliary optimization, \textbf{P4}, the theoretic performance guarantee can be further derived for the modified Lyapunov optimization method. This theorem demonstrates that the optimal objective value of the reformulated online problem, \textbf{P2}, has a gap from the optimal objective value of the original stochastic problem, \textbf{P1}. This performance gap, can be bounded by $\Theta$. The visual scheme of the proposed online distributed approach is illustrated in Figure \ref{apen_3}.

It should be noted that the bound for the performance gap, $\Theta$, can be further minimized in a distributed manner by optimizing the weight and shifting parameters of ESS. The detailed performance gap minimization algorithm, Algorithm \ref{apen_a2}, can be found in Section 6.9. Algorithm \ref{apen_a2} is performed at the offline stage to decide the optimal weight and shifting parameters before the start of the online operation period for each prosumer. Only one time of running is required for Algorithm \ref{apen_a2}. During the online operation, the initialized weight and shifting parameters are kept constant. The overall framework contains the performance gap minimization algorithm and the online distributed algorithm altogether.

\section{Numerical Simulations}
This section provides numerical analysis of the proposed online distributed algorithm and overall framework for the real-time P2P energy market operation. The algorithms are implemented in MATLAB 2020b and run on a windows 10 computer with 6-core Intel(R) i5-10500 CPU@3.10GHz and 16GB RAM. To illustrate the proper-ties and effectiveness of the proposed online distributed algorithm, one-day simulation with a one-hour time interval is carried out. Moreover, to demonstrate the performance of the proposed algorithm and overall framework, one-day simulation with a shorter 15-minute slot duration is carried out to emulate the real-time scenario.

\subsection{Simulation Setup}
\begin{figure}[htbp]
	\centering
	\includegraphics[width=1.0\linewidth]{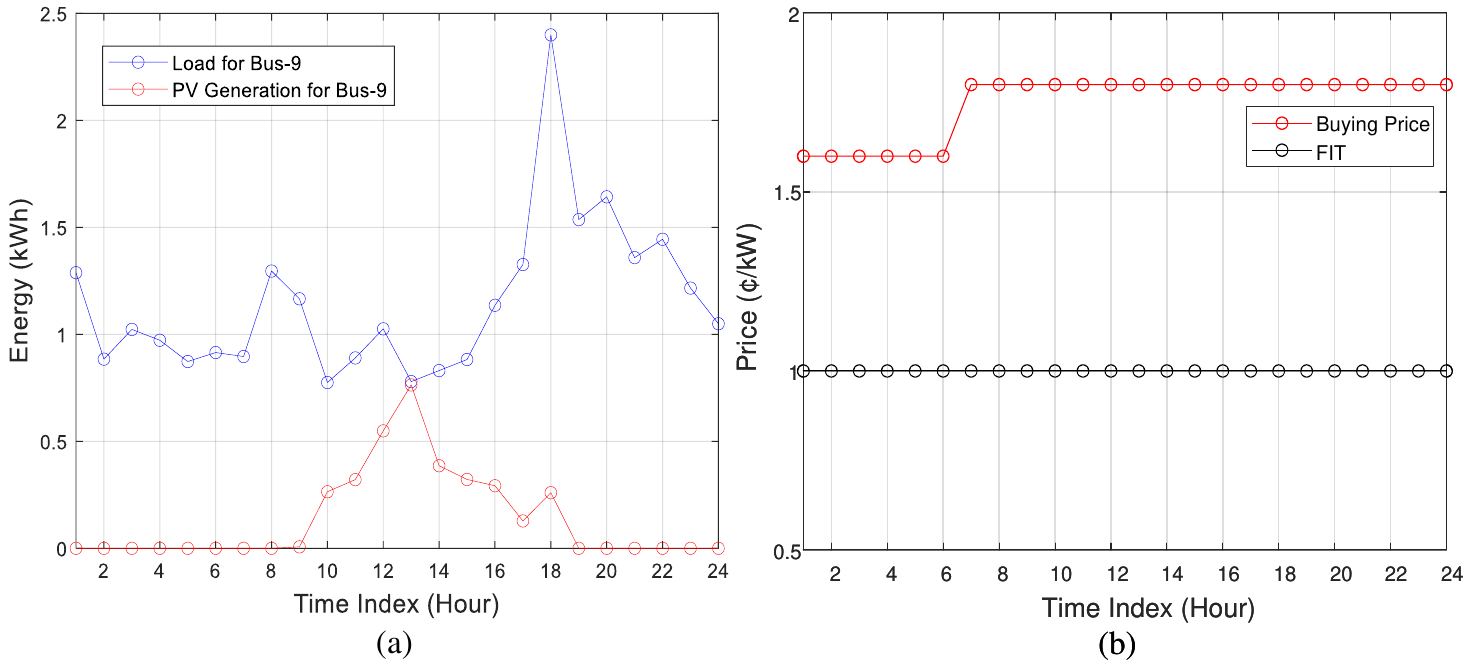}
	\caption{(a) Real PV generation and load data. (b) price signals from the utility company.}
	\label{apen_4}
\end{figure}

The time-of-use prices are considered in this paper. As shown in Figure \ref{apen_4}(b), buying prices from the utility company at the off-peak period (12:00 am-6.00 am) are assumed to be lower than those during the regular period. The standard IEEE 15-bus distribution system is employed to verify the proposed algorithm and framework, and the resistance and reactance data for this system are adopted from \cite{ouali2020improved}. These network data are considered to remain constant during the online operation period. Without loss of generality, the utility company is located at the substation bus 0, and the other buses are all equipped with the PV panels and energy storage systems. The one-day real PV power generation and load data are adopted for each prosumer from \cite{pena2019optimized}, as illustrated in Figure \ref{apen_4}(a).
The following parameters are randomly generated for different prosumers at each time slot: i) The parameters for the utility function of the buyers are randomly generated from the space $\alpha_i \in [0.02, 0.1]$¢/kWh$^2$, and $\beta_i \in [0.5, 2.0]$¢/kWh; ii) For the sellers, the corresponding parameters of the utility function are generated from the space $\alpha_i \in [0.1, 0.15]$¢/kWh$^2$, and $\beta_i \in [0.2, 0.6]$¢/kWh; iii) The parameter for the discomfort cost function for all the prosumers is generated from the space  $\gamma_i \in [2.5,3.5]$¢/kWh$^2$; iv) The parameter $S_i^{max}$ of ESS is generated from the space $[5,8]$kWh. For ESS, the parameter $S_i^{min}$ is set to be 0 and the parameter $\kappa_i$ is set to be 0.998.  For the network constraints, the minimum and maximum per-unit voltage magnitude at bus $i$ are set to be $V_i^{min}=0.95$(p.u.) and  $V_i^{max}=1.05$(p.u.). The minimum and maximum active power flows over the line $\ell$ are set to be $P_{\ell}^{min}=-7$kW and $P_{\ell}^{max}=7$kW. Meanwhile, the minimum and maximum reactive power flows over the line $\ell$ are set to be $Q_{\ell}^{min}=-7$(KVAR) and $Q_{\ell}^{max}=7$(KVAR). The limits of energy demand for prosumer $i$ are set to be $d_{i,t}^{min}=0.5\cdot d_{i,t}$ and $d_{i,t}^{max}=1.5\cdot d_{i,t}$. At the beginning of the online operation, $t$ is set to be 0, and $t^{max}$ is set to be the number of total time slots for the given operation period. $k^{max}$ is set to be 2000. Other parameters, such as $k,e_{i,j,t}^1,w_{i,t}^1,p{i,t}^1,u_{i,j,t}^1,h_{i,j,t}^1,\mu_{i,j,t}^1,\mathcal{R}_{i,t}^1,\boldsymbol{\bar{\rho}}^1,\boldsymbol{\underline{\rho}}^1$, are all initialized to be 0 at the beginning of each time slot. The proposed algorithm and overall framework are then verified over 7 cases.

\begin{figure}[!htbp]
	\centering
	\includegraphics[width=1.0\linewidth]{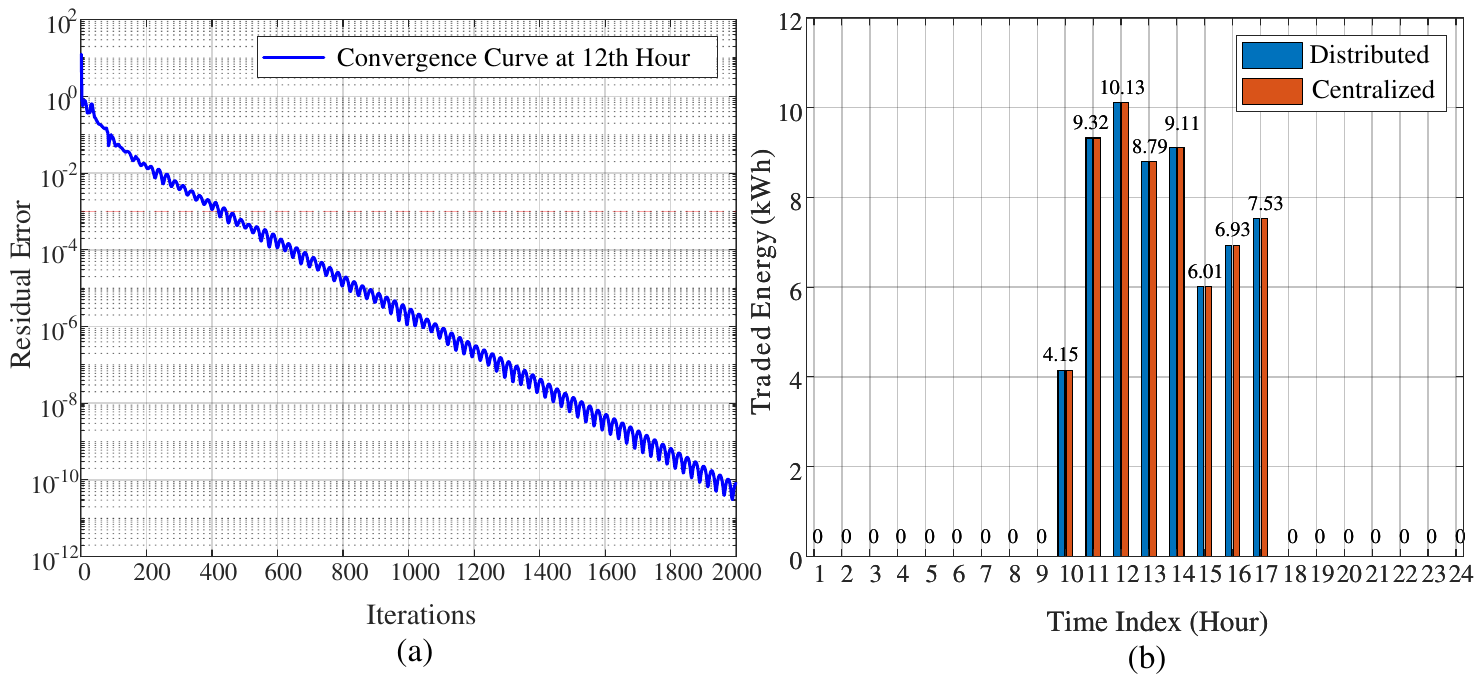}
	\caption{(a) The traded energy from two methods. (b) the convergence of the proposed algorithm.}
	\label{apen_5}
\end{figure}

\subsection{Optimality and Convergence of the Online Distributed Algorithm}
\begin{figure}[htbp]
	\centering
	\includegraphics[width=1.0\linewidth]{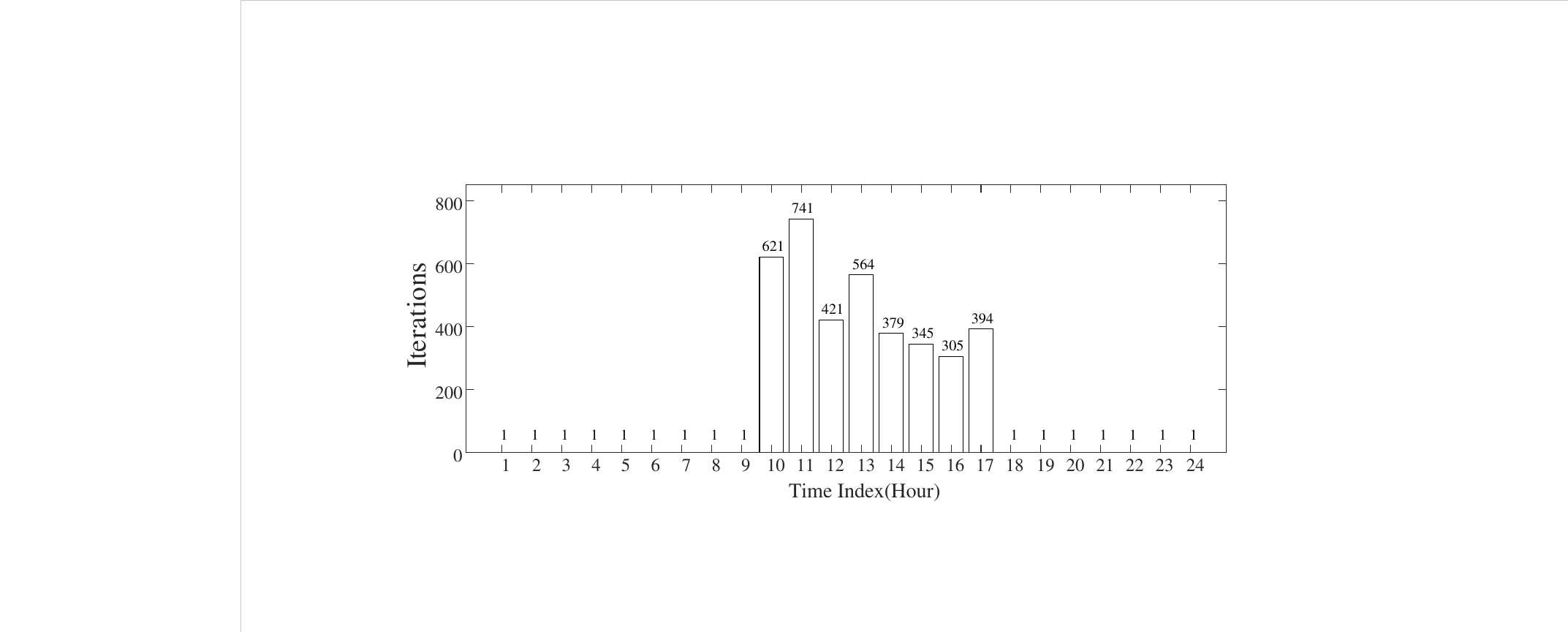}
	\caption{Iterations needed to converge at different time slots.}
	\label{apen_6}
\end{figure}

The proposed online algorithm is implemented in a distributed manner, and its optimality and computational efficiency are compared with the centralized method. The convergence criterion for the distributed method is set to be $10^{-3}$. During the time from the 10th hour to the 17th hour, there is surplus PV power generation. The P2P energy market operates normally during this period and the surplus energy can be traded among prosumers. For the centralized method, the reformulated optimization, \textbf{P2}, is solved. As illustrated in Figure \ref{apen_5}(b), centralized and distributed methods reach the same results of the total traded energy among prosumers for all the time slots. It demonstrates that the online distributed algorithm can obtain the same optimality as the centralized method. As for the convergence of the online distributed algorithm, which is shown in Figure \ref{apen_5}(a). At the 12th hour, the algorithm converges at around the 420th iteration to obtain the required accuracy. When the proposed algorithm runs about 2000 iterations, the residual reaches a relatively small value at an order of $10^{-10}$.
The iterations needed for the algorithm to converge to the satisfying accuracy are diverse at different time slots, as depicted in Figure \ref{apen_6}. During the time slots when the P2P energy market is not normally working, the required number of iterations is 1. When the P2P energy market is in operation, the required number of iterations to converge varies from 300 to 750, which is mainly due to the different ratio of buyers and sellers in the market and their distinct physical locations in the network. By comparing the execution time with the centralized method, the overall time expense of the distributed method during the one-day online operation period is significantly lower than the centralized method. 

\subsection{Effects of the Network Constraints}
\begin{figure}[htbp]
	\centering
	\includegraphics[width=1.0\linewidth]{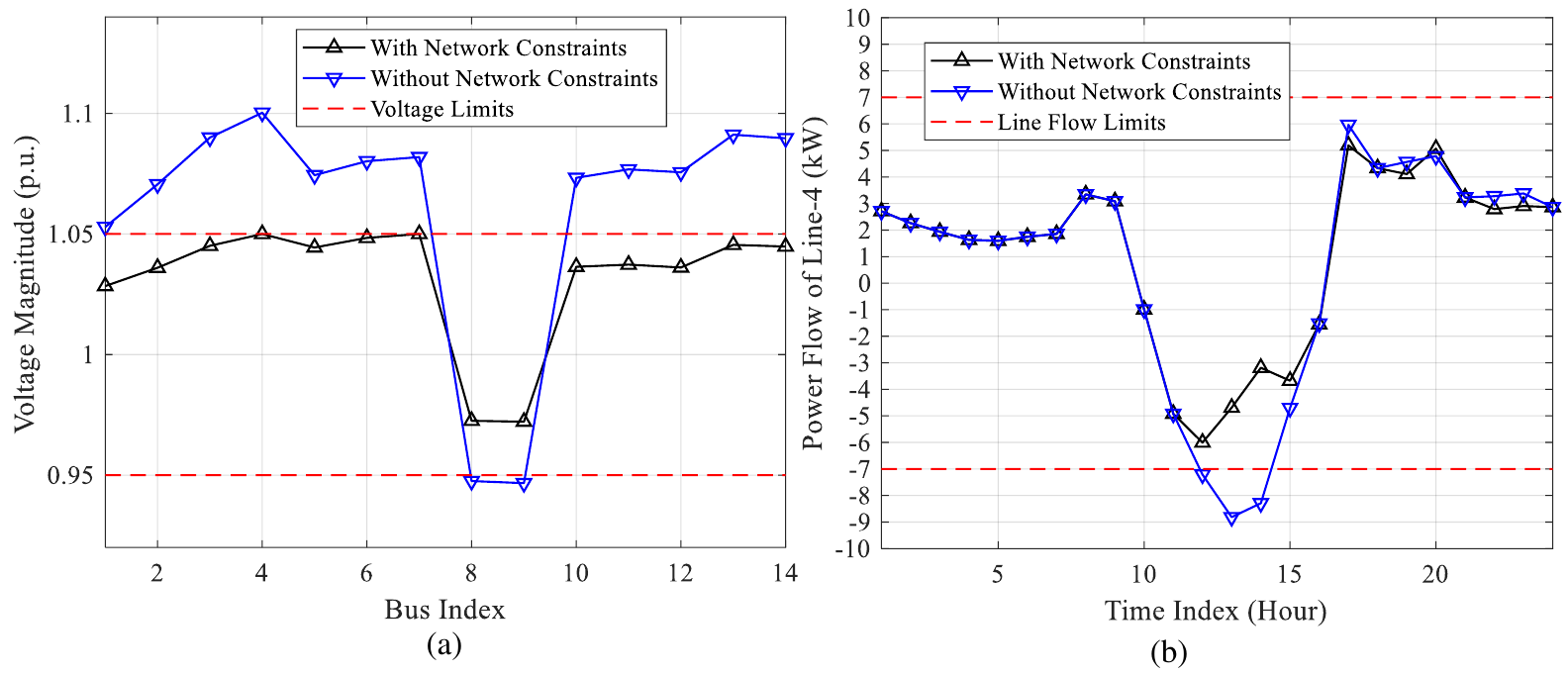}
	\caption{(a) The voltage magnitude profiles at 14th hour. (b) the power flow of line-4 overall time slots.}
	\label{apen_7}
\end{figure}

The physical network constraints are considered in this work, which help to secure the P2P energy trading. As shown in Figure \ref{apen_7}(a), the per-unit voltage magnitude profiles of all buses with the network constraints are com-pared with those without network constraints at the 14th hour. Without network constraints, certain energy trading transactions can be accepted while they might cause the bus voltage violations. The network constraints well help to regulate the per-unit voltage magnitude within the normal ranges. Likewise, the network constraints also help to regulate the power flow over the lines. As depicted in Figure \ref{apen_7}(b), without these constraints, the optimal power flow profiles are much more deviated. Comparably, with these constraints, some potential trading transactions are declined which might violate the network constraints during the period from the 10th to the 15th hour. The power flow of line-4 is therefore ensured to be within the normal range. In this way, it is crucial to consider network constraints when operating the real-time P2P energy market.

\subsection{Flexibility of and Effectiveness of the Online Distributed Algorithm}
This case analyses the flexibility and effectiveness of the proposed algorithm. The proposed online distributed algorithm is compared in regard of the flexibility of operating ESS with the state-of-the-art online algorithms: online regret algorithm \cite{guo2021online}, and traditional online algorithm \cite{shi2015real,zhong2019online}. As shown in Figure \ref{apen_9}(a), online regret algorithm can greatly constrain the actions of operating ESS due to the increasing impacts of the penalty term in the objective function with time going on. Moreover, for the traditional online algorithm, since the operating variable can only take either the boundary value or zero (e.g., -2, 0, and +2), the flexibility of operating ESS is also limited. Specifically, ESS charges energy during the normal operation period of P2P energy market, and then discharges energy for self-consumption when surplus energy from the prosumers is no longer available. 

\begin{figure}[htbp]
	\centering
	\includegraphics[width=1.0\linewidth]{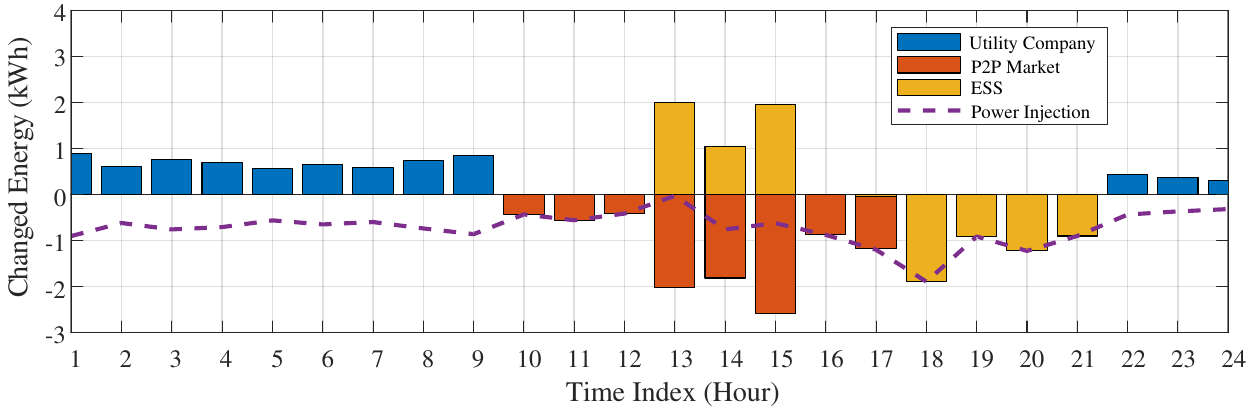}
	\caption{The energy profiles of bus-9.}
	\label{apen_8}
\end{figure}

It is observed that the charged or discharged energy of ESS at each time slot are within the allowed limits. The SoC of ESS is also within its normal range, which is from 0 to 5(kWh). This demonstrates the effectiveness of the proposed online distributed algorithm. As depicted in Figure \ref{apen_8}, detailed energy interactions are observed. For bus 9, the prosumer located on this bus is always a buyer in the market and the purple dashed line represents the prosumer’s net energy demands (active power injection). Before the normal operation of the P2P energy market, this prosumer feeds all its demands from the utility company. When there are sellers in the market, this prosumer can feed its demands from the sellers. When its demand is relatively lower, ESS is employed to store the energy from the P2P market. After the normal operation of P2P market, which is after the 17th hour, this prosumer can use its energy in ESS stored from the previous time slots to feed its demands. Thus, this prosumer no longer needs to purchase energy from the utility company.

\subsection{Performance of the Online Distributed Algorithm}
The performance of the online distributed algorithm and overall framework is analysed in this part. Five algorithms and one framework are considered here: (A1) Greedy Algorithm: This algorithm makes optimal decisions at each time slot and does not consider the future information, which can provide the worst upper bound. (A2) Traditional Online Algorithm: This is the traditional Lyapunov optimization method mentioned in \cite{shi2015real,zhong2019online}. The weight parameters $\delta_i$ and shifting parameter of prosumer $\epsilon_i$ takes some deviation from the optimal ones for demonstration, where $\delta_i=\delta_i^*+0.01$ and $\epsilon_i=\epsilon_i^*+1$, which keeps the same for algorithms A3 and A4. (A3) Online Regret Algorithm: This is the online optimization method mentioned in \cite{guo2021online}, and the step of minimizing the performance gap is also not included. (A4) Proposed Online Distributed Algorithm: This algorithm makes the optimal decisions at the current time slot and only relies on realizations of the past and current system information, and the step of minimizing the performance gap is also not included. (A5) Pro-posed Overall Framework: This framework contains A4 and the step of minimizing the performance gap, which provides the optimal weight parameters  $\delta_i^*$ and shifting parameter $\epsilon_i^*$ for prosumer $i$. (A6) Offline Algorithm: This algorithm makes optimal decisions over the whole time slots with all the system states are perfectly known, which can render the best lower bound.

\begin{figure}[htbp]
	\centering
	\includegraphics[width=1.0\linewidth]{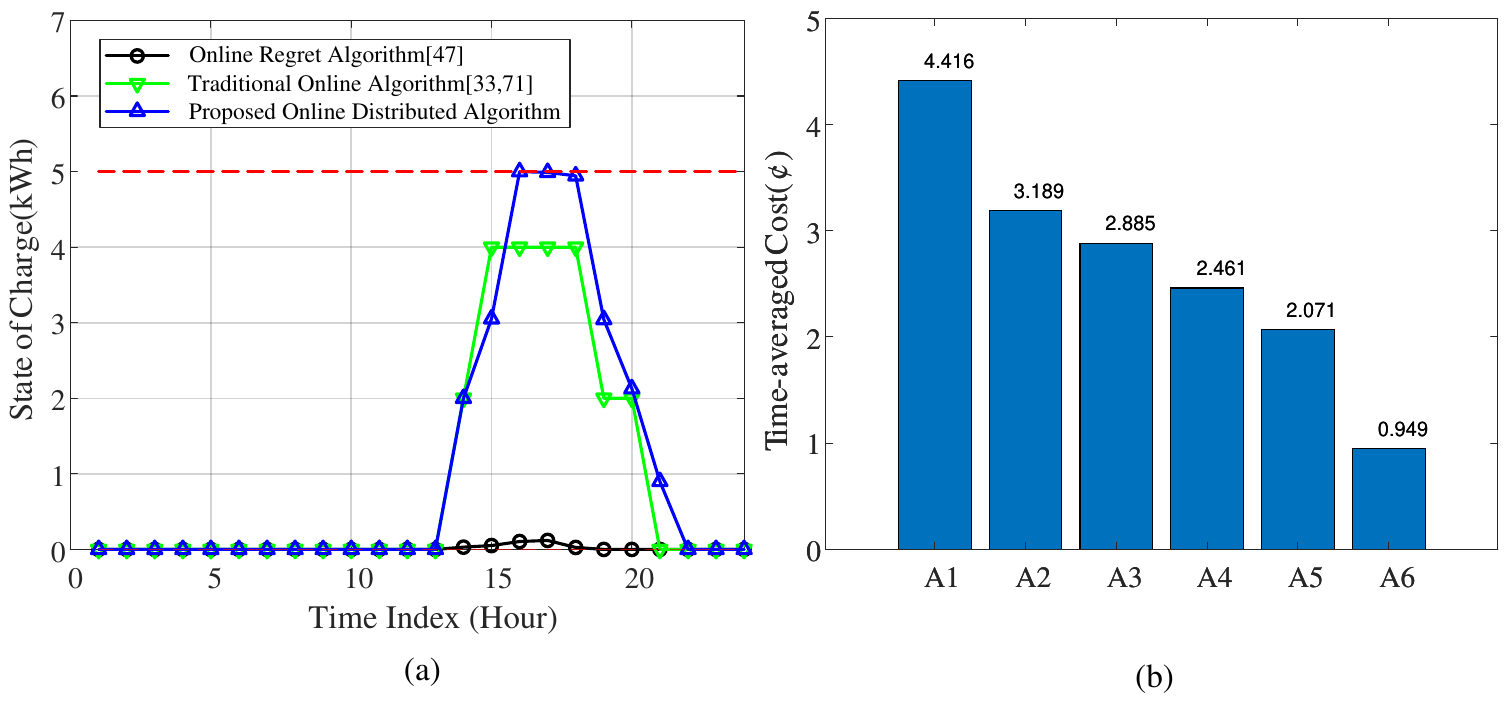}
	\caption{(a) Evolution of different optimization algorithms. (b) performance comparison.}
	\label{apen_9}
\end{figure}

The time-averaged costs obtained by the five algorithms and a framework are shown in Figure \ref{apen_9}(b). The overall framework brings approximately 15.85\% lower cost than the original proposed online distributed algorithm, 28.21\% lower cost than online regret algorithm, 35.06\% lower cost than the traditional Lyapunov algorithm, and 53.10\% cost reduction compared to the greedy algorithm. It demonstrates the proposed online distributed algorithm as well as the overall framework has better performance than the state-of-the-art online algorithms.

\subsection{Scalability of the Online Distributed Algorithm}
\begin{figure}[htbp]
	\centering
	\includegraphics[width=0.8\linewidth]{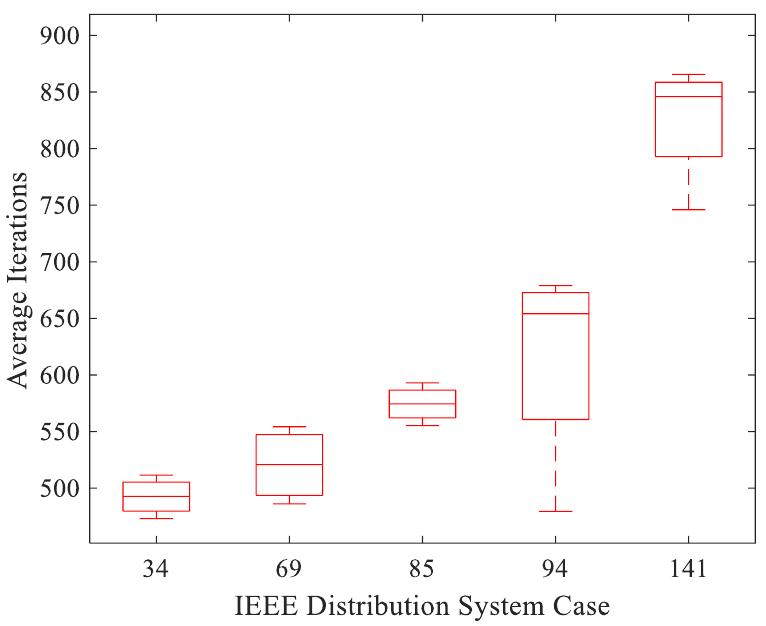}
	\caption{Scalability analysis of the proposed algorithm.}
	\label{apen_10}
\end{figure}

The scalability of the proposed online distributed algorithm is tested with different sizes of distribution bus systems ranging from 34 to 141. For the distribution network, the 141-bus system is considered to be large enough. Since the iteration needed to converge is 1 when the P2P energy market is not working, the averaged iteration only considers the iterations needed to converge when the P2P energy market is normally operated. It is ob-served from Figure \ref{apen_10} that when coming across a larger bus system, the proposed algorithm usually demands much more iterations to converge. The proposed online distributed algorithm is also tested under different penalty factors, $\eta$, ranging from 1 to 15. It is observed that the algorithm converges for all the scenarios. By finely tuning the penalty factor, the algorithm can converge much faster. Under all the scenarios, the average numbers of iterations are below 1000, which demonstrates its good scalability.
The real-time operation of the P2P energy market with a one-hour time interval is investigated. $\eta$ is fixed as 12. The online computation time refers to the aggregated computation time of 24 time slots for all the agents. For the offline stage, Algorithm \ref{apen_a2} is performed to optimize the system parameters for both centralized and distributed methods. Their computation time, as shown in Table \ref{apen_t2}, is almost the same. However, for the online stage, the proposed online distributed method, i.e., Algorithm \ref{apen_a1}, has a higher computational efficiency com-pared to the traditional centralized method. As shown in Table \ref{apen_t3}, for the 15-bus system, the online distributed method is approximately 100 times faster than the centralized method if communication costs are ignored. When the system size increases to 141, the online distributed method can be about 450 times faster than the centralized method. The reason for the significant differences is that a solver must be used to solve the centralized problem. In this work, CVX toolbox with IPOPT is employed as the solver \cite{grant2014cvx}. Comparably, it is more intuitive and faster to iteratively update the variables through closed-form solutions for the decomposed sub-problems. Thus, the proposed online distributed algorithm is more appropriate for the real-time P2P energy market operation. 

\begin{table}[htbp]
	\centering
	\caption{Offline computation time of two methods (s)}
	\label{apen_t2}
	\resizebox{1.0\columnwidth}{!}{
	\begin{tabular}{ccccccc}
		\toprule  
		& 15-bus &34-bus & 69-bus &85-bus & 94-bus &141-bus \\ 
		\cmidrule(r){2-7}
		{Centralized Method}&1.04 &2.45&5.16 &6.07 &6.71 &10.04 \\		
		{Online Distributed Method}&1.03&2.39&5.27 &6.87 &7.21 &10.16\\
		\bottomrule 
	\end{tabular}}
	\vspace{-0.5cm}
\end{table}\

\begin{table}[htbp]
	\centering
	\caption{Online computation time of two methods (s)}
	\label{apen_t3}
	\resizebox{1.0\columnwidth}{!}{
	\begin{tabular}{ccccccc}
		\toprule  
		& 15-bus &34-bus & 69-bus &85-bus & 94-bus &141-bus \\ 
		\cmidrule(r){2-7}
		{Centralized Method}&19.09 &69.83&334.32 &878.86 &1617.14 &5658.53 \\		
		{Online Distributed Method}&0.16&0.73&3.21 &4.91 &3.97 &12.40\\
		\bottomrule 
	\end{tabular}}
	\vspace{-0.5cm}
\end{table}\

\subsection{Sensitivity Analysis of the Proposed Online Distributed Algorithm}
\begin{figure*}[htbp]
	\centering
	\includegraphics[width=0.8\linewidth]{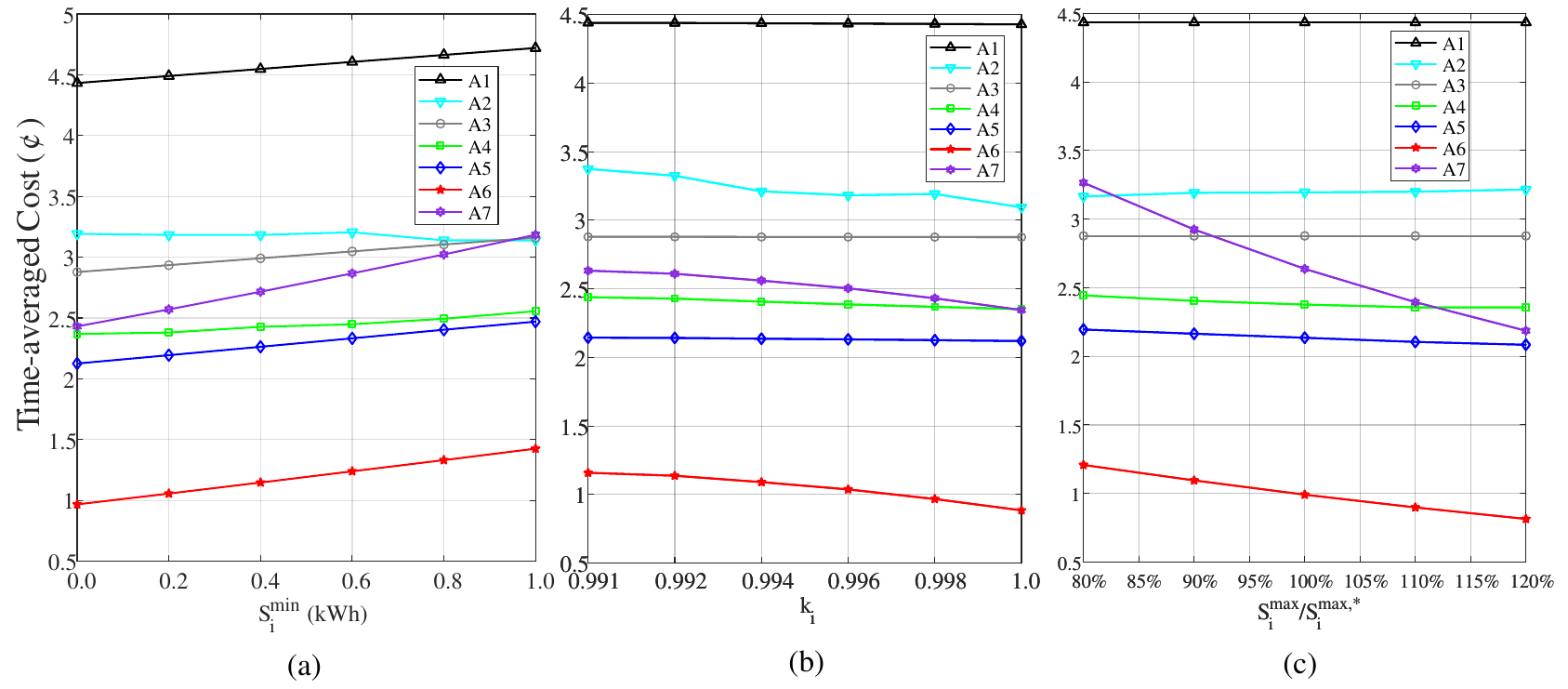}
	\caption{Time-averaged costs under different scenarios: (a) minimum values of SoC. (b) maximum values of SoC. (c) charging/discharging coefficient.}
	\label{apen_11}
	\vspace{-0.5cm}
\end{figure*}

The sensitivity analysis is carried out to demonstrate the performances of algorithms under different settings. The theoretical performance guarantee is included as A7. (A7) Theoretical Guarantee: The guarantee is denoted as $\Psi_{P1}^*(\boldsymbol{\Phi})+\Theta$, where $\Psi_{P1}^*(\boldsymbol{\Phi})$ is the optimal objective value obtained by A6. $S_i^{max}$ is tested in the range of 1$\pm$20\% of the given nominal value $S_i^{max,*}$ for all the agents based on \cite{zhong2019online,huang2017market}. $S_i^{min}$ is set to be in the range of 0$\sim$10\% of $S_i^{max}$, which is a common practice to ensure the normal operation of ESS \cite{conover2014protocol}. The charging/discharging coefficient $\kappa_i$ is set to be the realistic range $\kappa_i \in (0.99,1]$ as in \cite{zhong2019online}. As in Figure \ref{apen_11}(a), the increase of $S_i^{min}$ will bring more costs due the decreased capacity of ESS, since the ability of ESS is limited to store electricity when the price is less expensive. Similarly, as shown in Figure \ref{apen_11}(b), the increase of $S_i^{max}$ will lead to less costs due to the increased capacity of ESS. As shown in Figure \ref{apen_11}(c), the larger charging/discharging coefficient causes less costs since the energy loss of operating ESS is being ignored, and thus the capacity of ESS will be less influenced after frequent operations. Moreover, the proposed algorithm obtains better performance than the other online algorithms under different situations. Allied with the relaxation minimization step, the proposed algorithm can obtain even much lower costs as denoted by A5. The theoretical guarantee, as denoted by A7, keeps the same trends with the proposed algorithm, framework, and the offline algorithm. The theoretical guarantee can be functioned as the performance indicator for the proposed framework, and it can be tighter if none-i.i.d. situations of the system states are considered. Overall, the proposed framework obtains the least costs than the state-of-art online algorithms.

\subsection{Model Extensions}
1) Cost Allocation of Power Losses: The proposed overall framework can also cover the scenario in which power losses are considered. In order to motivate prosumers to trade with the nearest neighbours, the cost of power losses is further considered and allocated \cite{ullah2021peer}. The cost is related to the electrical distance and amount of traded energy \cite{yang2021proof}. Let $\tau$ be the cost coefficient for delivering per square unit of traded energy over per unit of electrical distance \cite{paudel2020peer,baroche2019exogenous}. $Z_{\ell}$ denotes the impedance of line segment $\ell$, and $L_i^j$ is the set containing lines segments connecting prosumers $i$ and $j$. $g_{i,j}$ denotes the cost for delivering per square unit of traded energy be-tween prosumers $i$ and $j$. The cost for prosumer $i$, $\mathcal{C}_{i,t}$, at time slot $t$ is formulated as follows:
\begin{subequations}
	\begin{align}	
		&g_{i,j}=\tau \cdot \sum_{\ell \in L_i^j}|Z_{\ell}|  \label{apen_f29}\\
		&\mathcal{C}_{i,t}=\sum_{j \in \mathcal{N}_i} g_{i,j}e_{i,j,t}^2 . \label{apen_f30}
	\end{align}
\end{subequations}

\begin{figure}[htbp]
	\centering
	\includegraphics[width=1.0\linewidth]{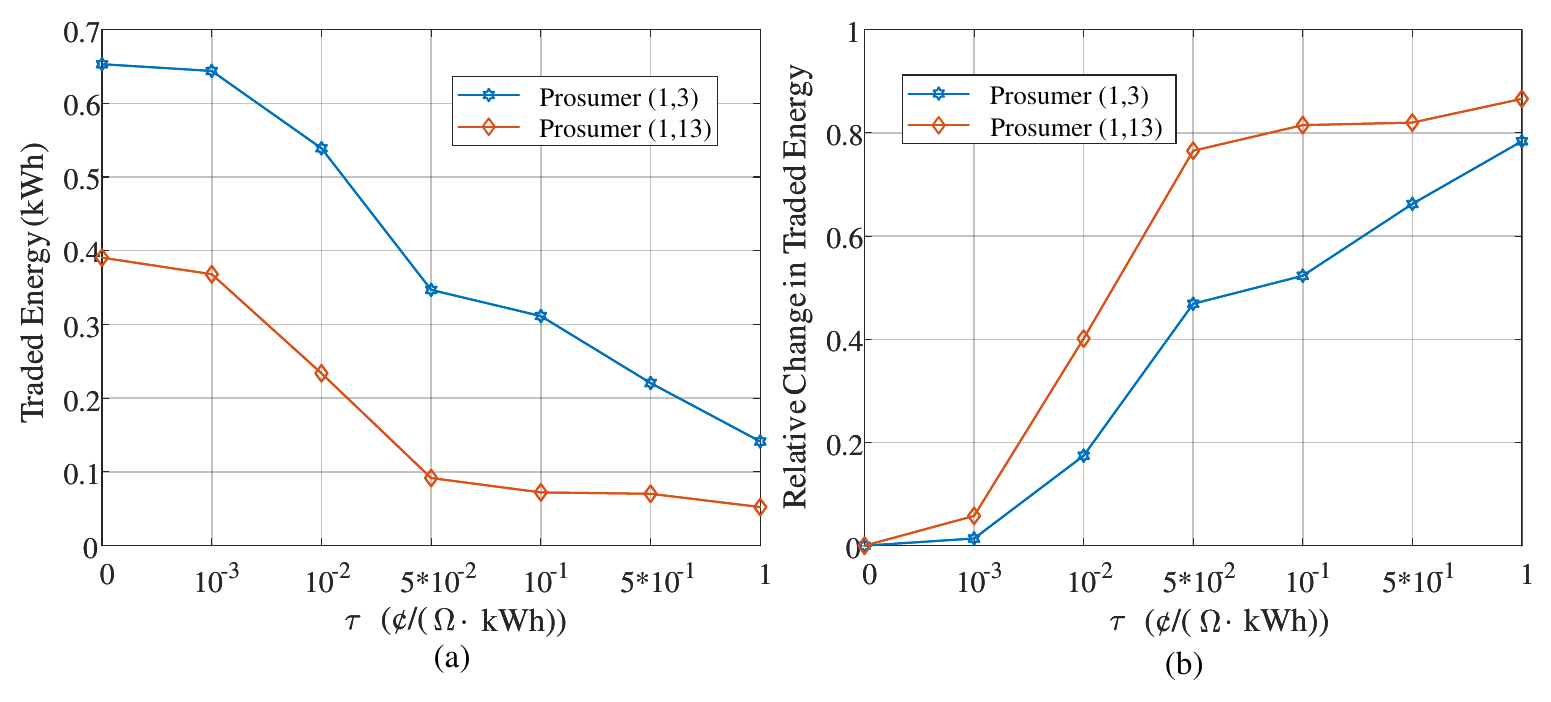}
	\caption{(a) Traded energy with different $\tau$. (b) relative change of traded energy with different $\tau$.}
	\label{apen_12}
\end{figure}

$\mathcal{C}_{i,t}$ is an additional term to be added to the objective function \eqref{apen_f20} when power losses are considered. The inclusion of \eqref{apen_f29} and \eqref{apen_f30} does not change the structure of the proposed framework. As shown in Figure \ref{apen_12}(a), the inclusion of power losses leads to less traded energy between prosumers in the P2P energy market, and larger cost coefficient has stronger influences on the energy transactions. As delineated in Figure \ref{apen_12}(b), the inclusion of power losses has greater impacts on the amount of traded energy between prosumers 1 and 13 than that be-tween prosumers 1 and 3. In the real power network \cite{ouali2020improved}, prosumer 1 has a shorter electrical distance from prosumer 3 compared to that between prosumers 1 and 13. The test results indicate that the inclusion of power losses contributes to the decrease of traded energy, and prosumers are more liable to trading energy with the nearest prosumers.

\subsection{Performance Gap Minimization}
To make the original problem \textbf{P1} available for online optimization, it is reformulated to \textbf{P2} and \textbf{P4}. However, the performance gap $\Theta$ remains between \textbf{P1} and \textbf{P2}. It is observed that the performance gap $\Theta$ is only related to ESS system parameters. To minimize the performance gap, this part first formulates the minimization problem \textbf{P5}. \textbf{P5} is then reformulated into a Semidefinite Program (SDP), \textbf{P6}, which is solved by the off-shelve solvers to minimize the performance gap $\Theta$.

1) Sensitivity Analysis of $\delta_i$: To obtain the appropriate range for the weight parameter $\delta_i$ and provide the efficient bound $\Theta$ for the proposed online distributed method, the sensitivity analysis is conducted with regard to $\delta_i$. Boundary constraints (11) and (12) for ESS also work for the reformulated problems, which always ensure the norm operation of ESS under any shifting parameters. Therefore, a loose boundary can be rendered for the shifting parameter $\epsilon_i$. Since $\epsilon_i$ always take the negative value \cite{neely2022stochastic}, it can be reasonably bounded by the nega-tive values of the minimum and maximum status of ESS. The initial bounds for $\delta_i$ and $\epsilon_i$ are given as follows:
\begin{subequations}
	\begin{align}	
		&\delta_i^{min} \le \delta_i \le \delta_i^{max} \label{apen_f23}\\
		&S_i^{max} \le \epsilon_i \le -S_i^{max}. \label{apen_f24}
	\end{align}
\end{subequations}
To conduct the sensitivity analysis for $\delta_i$, we set $\delta_i^{min}=\delta_i^{max} \in  \{10^{-5},10^{-4},10^{-3},10^{-2},5\times 10^{-5},10^{-1}\}$. Algorithm \ref{apen_a2} is performed to automatically choose the appropriate shifting parameter $\epsilon_i$.
\begin{figure}[htbp]
	\centering
	\includegraphics[width=1.0\linewidth]{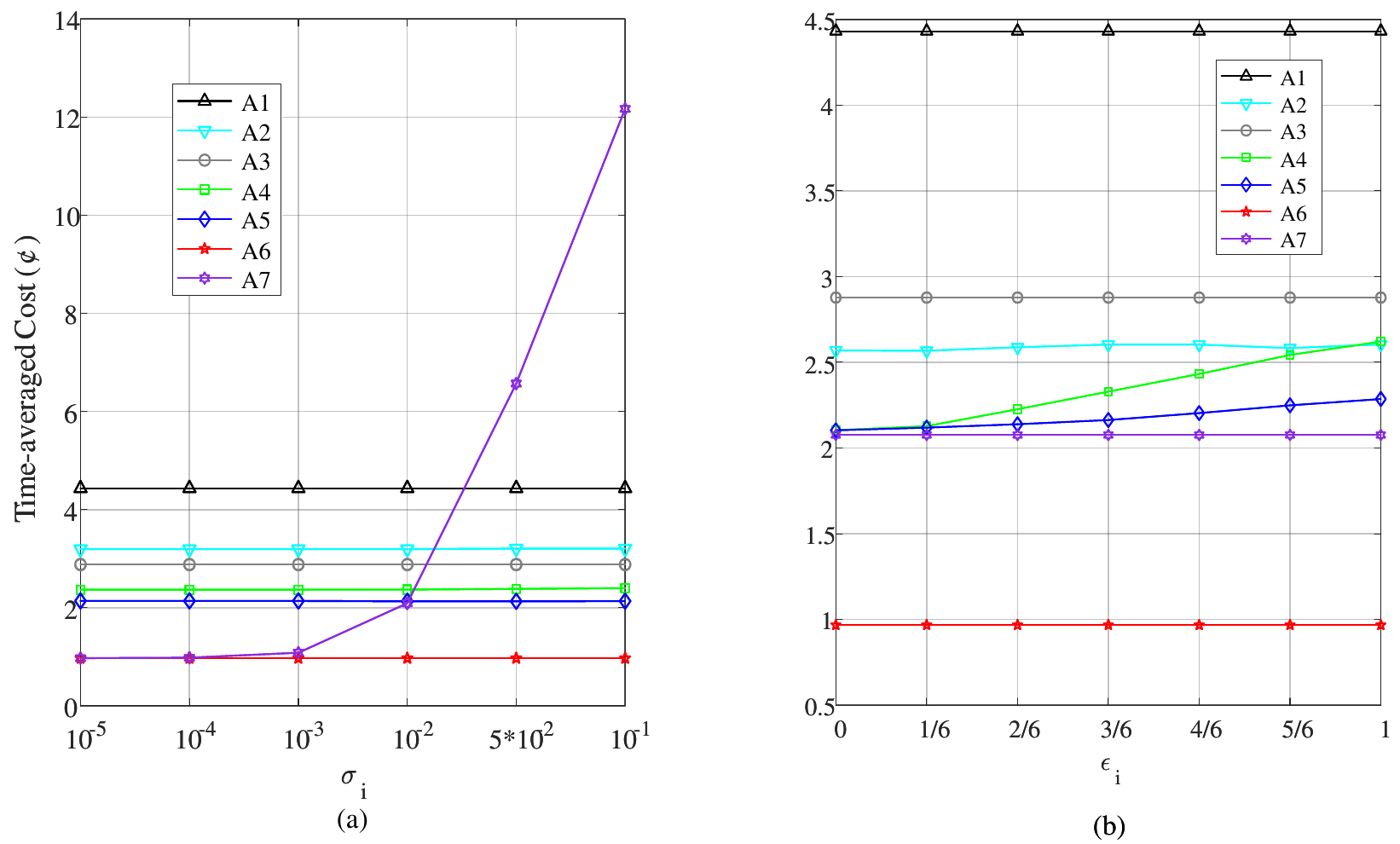}
	\caption{(a) Analysis of the weight parameter $\delta_i$. (b) analysis of the shifting parameter $\epsilon_i$.}
	\label{apen_13}
\end{figure}

Testing results from Figure \ref{apen_13}(a) show that the choice of $\delta_i$ has little impacts on the practical performance of the proposed algorithm A4 and framework A5. However, A5 always performs better than A4, and this is due to the optimal value of $\epsilon_i$ obtained by Algorithm \ref{apen_a2}. However, the choice of $\delta_i$ does have impacts on the theoretical performance bound $Theta$. Larger $\delta_i$ will result in larger $Theta$, and this result is in consistent with the existing findings \cite{shi2022lyapunov}. To provide an efficient performance bound $Theta$, we refer to \cite{shi2022lyapunov} and set $\delta_i^{min}=10^{-2}$ and $\delta_i^{max}=5\times10^{-2}$. It should be noted that for other scenarios with different lengths of the operation period, a larger and rougher range for the weight parameter $\delta_i$ can be applied.

2) Sensitivity Analysis of $\epsilon_i$: The impacts of shifting parameter $\epsilon_i$ on the performances of different algorithms are investigated. $\epsilon_i$ is kept in the range of $[10^{-2},5\times10^{-2}]$. $\epsilon_i$ is set to be $\epsilon_i=\epsilon_i^{min}=\epsilon_i^{max}=-S_i^{max}+\mathcal{X}_i(-S_i^{min}+S_i^{max})$ with $\mathcal{X}_i \in \{0,1/6,2/6,3/6,4/6,5/6,1\}$. For A4, $\delta_i$ is fixed to be $4\times10^{-2}$. For A5, Algorithm \ref{apen_a2} is conducted to obtain the optimal $\delta_i$. As shown in Figure \ref{apen_13}(b), larger $\epsilon_i$ introduces worse performance for the proposed algorithm A4 and framework A5. A5 still performs better than A4 and other online methods A2 and A3. The choice of $\epsilon_i$ has little influence on the theoretical guarantee. 

3) Performance Gap Minimization: The performance gap minimization problem is separable to each prosumer and is defined for prosumer i as below:
\begin{subequations}
	\begin{align}	
		\textbf{P5}: & \qquad \mathop{min}_{\delta_i,\epsilon_i} \quad \Pi_i+\frac{1}{2}\delta_i\cdot(\kappa_i-1)^2\cdot(S_i^{max})^2+\frac{1}{2}\delta_i\cdot(\kappa_i-1)^2\cdot\epsilon_i^2 \notag\\
		&\qquad\qquad\qquad+\frac{1}{2}\delta_i\cdot(w_i^{max}-w_i^{min})^2\\
		&\qquad s.t. \quad \eqref{apen_f23},\eqref{apen_f24},
	\end{align}
\end{subequations}
where
\begin{align}	
	\Pi_i=\delta_i\cdot(\kappa_i-\frac{1}{2}-\frac{1}{2}\kappa_i^2)\cdot max\{(S_i^{min}+\epsilon_i)^2,(S_i^{max}+\epsilon_i)^2\}.  \label{apen_f25}
\end{align}
\textbf{P5} is in a complicated form, and it is reformulated into \textbf{P6} by applying the Schur Complement \cite{qin2014modeling,zhong2019online} to the constraint \eqref{apen_f25}:
\begin{subequations}
	\begin{align}	
		\textbf{P6}: \qquad &\delta_i\cdot(\kappa_i-\frac{1}{2}-\frac{1}{2}\kappa_i^2)\cdot \Lambda_i  +\frac{1}{2}\delta_i\cdot(\kappa_i-1)^2\cdot(S_i^{max})^2 \notag\\
		&\qquad+\frac{1}{2}\delta_i\cdot(\kappa_i-1)^2\cdot\epsilon_i^2+\frac{1}{2}\delta_i\cdot(w_i^{max}-w_i^{min})^2\\
		&\qquad s.t. \quad \eqref{apen_f23},\eqref{apen_f24},
	\end{align}
\end{subequations}
where
\begin{subequations}
	\begin{align}
		\left[\begin{array}{cc} \Lambda_i &(S_i^{min}+\epsilon_i) \\(S_i^{min}+\epsilon_i) &1  \end{array}\right] \succeq 0 \\
		\left[\begin{array}{cc} \Lambda_i &(S_i^{max}+\epsilon_i) \\(S_i^{max}+\epsilon_i) &1  \end{array}\right] \succeq 0.
	\end{align}	
\end{subequations}
It is noted that by setting the weight parameter to be a constant, \textbf{P6} becomes a SDP that can be solved by the off-shelf solvers. Therefore, Algorithm \ref{apen_a2} is designed to solve \textbf{P6} and only one time of running is required.

\begin{algorithm}[!htbp]
	\caption{Performance Gap Minimization Algorithm.}\label{apen_a2}
	\begin{algorithmic}
		\STATE 
		\STATE {\textbf{Initialization:}}
		\STATE \hspace{0.5cm}$ \text{Initialize } \delta_l \leftarrow \delta_i^{min}, \delta_r \leftarrow \delta_i^{max}, O^*, \delta_\epsilon$ 
		\STATE {\textbf{Algorithm for each prosumer} $i \in \mathcal{N}_p$}
		\STATE \hspace{0.5cm}$\textbf{while } \delta_l \le \delta_r \textbf{ do}$ 
		\STATE \hspace{0.8cm}$\text{Set } \delta_l \leftarrow \delta_l+\delta_\epsilon;$
		\STATE \hspace{0.8cm}$\text{Solve \bf{P6} with } \delta_l, \text{and obtain optimal  } \hat{\epsilon}_i, \hat{O}_i;$
		\STATE \hspace{0.8cm}$\textbf{If } [\hat{O}_i]^+ \le O_i^* \textbf{ then}$	
		\STATE \hspace{1.5cm}$\text{Set } O_i^* \leftarrow \hat{O}_i, \epsilon_i^* \leftarrow \hat{\epsilon_i}, {\delta_i}^* \leftarrow \delta_l;$	  
		\STATE \hspace{0.8cm}$\textbf{end }$
		\STATE \hspace{0.5cm}$\textbf{end }$ 
	\end{algorithmic}
\end{algorithm}
\section{Conclusions}
Large penetration of renewable energy into the power system can shift the traditional centralized energy market to a distributed P2P energy market, which is promising in the future. This paper investigates the real-time energy trading in the P2P energy market under the physical network constraints when prosumers are equipped with PV panels and ESS. Since ESS is a time-coupling component in the power system and system states are highly uncertain in real situations, it makes the P2P energy market hard for online operation. If the optimal decisions for prosumers in the P2P energy market were made at each time slot, the worst performance can be obtained. Conversely, the future system states are hard to be accurately predicted especially for a long operation period, challenging the data-driven methods. Different from them, we formulate the real-time P2P energy trading problem as a spatial-temporally constrained stochastic optimization problem and then propose a light-weight method based on the modified Lyapunov optimization to approximately reformulate the stochastic optimization problem into an online one. The proposed method can provide better flexibility and performance for the real-time P2P energy market operation compared with the state-of-the-art online methods. Meanwhile, to protect the prosumers’ privacy, an online distributed algorithm based on the consensus ADMM is developed to solve the reformulated online problem. The closed-form solutions for all the sub-problems are derived, which helps to increase the computational efficiency and ideally brings approximately 100 times faster speed compared with the centralized method. Moreover, the proposed online distributed algorithm also owns good scalability, which can converge to the required accuracy within 1000 iterations under different scenarios. The theoretical near-optimal performance bound for the proposed online distributed algorithm is derived, and a performance gap minimization algorithm can be further applied to improve its performance. The simulation results show that the proposed online distributed algorithm can ensure the fast, safe, and stable long-term online operation of a real-time P2P energy market. The overall framework can bring approximately 15.85\% lower cost than the original proposed online algorithm, 28.21\% lower cost than the online regret algorithm, 35.06\% lower cost than the traditional Lyapunov algorithm, and a 53.10\% cost reduction compared to the greedy algorithm for a one-day operation.

\section*{Acknowledgements}
This work was supported in part by the joint Research Fund in Smart Grid (Grant No. U1966601) under cooperative agreement between the National Natural Science Foundation of China (NSFC) and State Grid Corporation of China (SGCC), and in part by the Research Grants Council of Hong Kong under Grant GRF 17209419.
\bibliographystyle{elsarticle-num}
\bibliography{example}
\end{document}